\documentclass[%
 reprint,
superscriptaddress,
nofootinbib,
 amsmath,amssymb,
 aps,
]{revtex4-2}
\usepackage{multirow}
\usepackage{subfigure}
\usepackage{graphicx}
\usepackage{dcolumn}
\usepackage{bm}
\usepackage[colorlinks,
            linkcolor=red,
            anchorcolor=blue,
            citecolor=blue
            ]{hyperref}

\graphicspath{ {figure/} }
\begin{document}

\title{Can the Production Cross-Section Uncertainties Explain the Cosmic Fluorine Anomaly?}

\author{Meng-Jie Zhao}
\email{zhaomj@ihep.ac.cn}
 \affiliation{%
 Key Laboratory of Particle Astrophysics, Institute of High Energy Physics, Chinese Academy of Sciences, Beijing 100049, China}
\affiliation{
 University of Chinese Academy of Sciences, Beijing 100049, China 
}%
 \author{Xiao-Jun Bi}
 \email{bixj@mail.ihep.ac.cn}
\affiliation{%
 Key Laboratory of Particle Astrophysics, Institute of High Energy Physics, Chinese Academy of Sciences, Beijing 100049, China}
\affiliation{
 University of Chinese Academy of Sciences, Beijing 100049, China 
}%
\author{Kun Fang}
\email{fangkun@ihep.ac.cn}
\affiliation{%
 Key Laboratory of Particle Astrophysics, Institute of High Energy Physics, Chinese Academy of Sciences, Beijing 100049, China}



\date{\today}

\begin{abstract}
The stable secondary-to-primary flux ratios of cosmic rays (CRs), represented by the boron-to-carbon ratio (B/C), are the main probes of the Galactic CR propagation. However, the fluorine-to-silicon ratio (F/Si) predicted by the CR diffusion coefficient inferred from B/C is significantly higher than the latest measurement of AMS-02. This anomaly is commonly attributed to the uncertainties of the F production cross sections. In this work, we give a careful test to this interpretation. We consider four different cross-section parametric models. Each model is constrained by the latest cross-section data. We perform combined fits to the B/C, F/Si, and cross-section data with the same propagation framework. Two of the cross-section models have good overall goodness of fit with $\chi^2/n_{d.o.f.}\sim1$. However, the goodness of fit of the cross-section part is poor with $\chi^2_{\rm{cs}}/n_{\rm{cs}}\gtrsim2$ for these models. The best-fitted F production cross sections are systematically larger than the measurements, while the fitted cross sections for B production are systematically lower than the measurements. This indicates that the F anomaly can hardly be interpreted by neither the random errors of the cross-section measurements nor the differences between the existing cross-section models. We then propose that the spatially dependent diffusion model could help to explain B/C and F/Si consistently. In this model, the average diffusion coefficient of the Ne-Si group is expected to be larger than that of the C-O group.
\end{abstract}
\maketitle


\section{\label{sec:level1}INTRODUCTION}
The stable secondary-to-primary flux ratios of cosmic rays (CRs) are essential probes for the Galactic CR propagation. The boron-to-carbon ratio (B/C) is considered the most effective one, as B is entirely secondary (mainly from C, N, and O) and has better-known production cross sections than Li and Be \cite{Strong:2007nh}. Assuming spatially independent CR propagation and continuously distributed CR sources, the propagation parameters inferred from B/C should be applicable to the calculation of other secondary-to-primary ratios. For example, the antiproton-to-proton ratio ($\bar{p}/p$) can be reasonably interpreted in the same framework, although a slight excess appears at $\sim$10~GeV of the AMS-02 spectrum \cite{AMS:2016oqu}, which is attributed to a dark matter signal \cite{Cui:2016ppb} or the fragmentation cross-section uncertainties \cite{Luque:2021ddh}. However, the Li spectrum predicted by the parameter sets inferred from B/C are significantly lower than the AMS-02 measurements \cite{Aguilar:2018njt} in a broad energy range, which may be attributed to the fragmentation cross-section uncertainties \cite{DelaTorreLuque:2021ayc,2022arXiv220300522M} or primary sources of Li \cite{Kawanaka:2017cae,Boschini:2019gow}.

The rigidity dependence of primary CRs Ne, Mg, and Si is found to be different from He, C, and O \cite{AMS:2020cai}. This indicates that these heavy primary CRs could belong to a distinctive class from the light primary CRs. Fluorine (F) is one of the main fragments of Ne, Mg, and Si, and its energy spectrum has been accurately measured by AMS-02 \cite{AMS:2021tnd}. As F is almost purely secondary \cite{Boschini:2020jty}, the secondary-to-primary flux ratio F/Si can probe the propagation of heavier CRs compared with B/C. However, the measured rigidity dependence of F/Si is distinctly different from B/C (or B/O), and F/Si calculated by the parameter sets inferred from B/C is significantly larger than the AMS-02 data (see Fig.~3 in \cite{AMS:2021tnd}). Unlike the case of Li, this overestimation cannot be interpreted by assuming primary sources of F.

There have been several works focusing on the F anomaly, which all attribute the F anomaly to the insufficiently constrained errors in the production cross sections \cite{Boschini:2021ekl,DelaTorreLuque:2021ayc,Vecchi:2021qca}. Although these works adopted different tools for solving the CR propagation, their results all indicated that the renormalization of $10\%-20\%$ on the production cross sections of F is required to interpret the discrepancy between the prediction and the data. A more recent work \cite{2022arXiv220801337F} performed a deeper analysis taking several systematic effects into account, including the energy correlations for the F/Si AMS-02 data, the systematic errors of the nuclear production cross sections, and the solar modulation uncertainties. They found that a $10\%$ overall increase for the production cross sections of B and $5\%$ overall decrease for that of F are needed to fit the data best. They argued that the propagation parameters obtained from F/Si data are compatible with those obtained from the AMS-02 (Li, Be, B)/C data, as the required renormalizations are within the typical range of the cross-section uncertainties.

In this work, we perform a combined analysis of F/Si, B/C, and the corresponding cross-section measurements to test whether the secondary-to-primary ratios for light and heavy nuclei can be naturally interpreted by a uniform propagation framework. The {\footnotesize GALPROP} code is used to solve the propagation equations. Our test has two main differences from \cite{2022arXiv220801337F}. First, we adopt several different parametrizations (or parametric formulas) for the production cross sections to discuss their systematic uncertainties. The normalization of each parametrization is constrained by the latest cross-section measurements. By comparison, \cite{2022arXiv220801337F} also modified different parametrizations to fit the updated cross-section data, but they estimated the uncertainty by the differences among the \textit{original} parametrizations. It means that the allowed cross-section variances are broader than the data constraints, and the uncertainties could be overestimated. 
Second, we do not use a single proxy channel for F or B to calculate the global renormalization for the production cross-section uncertainties.
Instead, the normalization of each reaction channel is separately constrained by the corresponding data (if the data is available). This could be the more reasonable approach as the energy dependence of the influence on the total flux varies with the reaction channel.

This paper is organized as follows. In Sec.~\ref{sec:level2}, we introduce the CR propagation model and the cross-section datasets used for analysis. In Sec.~\ref{sec:result}, we first determine the Gaussian-distributed normalization of each reaction channel by fitting the cross-section data as the preliminary step. Then we perform combined fits to the secondary-to-primary ratios measured by AMS-02 and the data-driven normalizations of the cross sections obtained in the preliminary step. The fitting result shows that the nuclei ratio and the cross-section data cannot be consistently explained even by trying different parametrizations of the cross sections. We propose a new scenario in Sec.~\ref{sec:effective}, in which the spatially dependent propagation accounts for the F anomaly. Sec.~\ref{sec:conclusion} is the summary of our findings above.

\section{\label{sec:level2}CALCULATION SETUP}
\subsection{\label{subsec:proper}Propagation}
We adopt the standard CR propagation model with reacceleration. 
Generally, the propagation equation of galactic CRs is expressed by
\begin{eqnarray}
	 {\frac{\partial \psi}{\partial t}}=&&q(x,p)+\nabla\cdot(D_{xx}\nabla\psi-V_c\psi)
	 +{\frac{\partial}{\partial p}}[p^2D_{pp}{\frac{\partial}{\partial p}}({\frac{\psi}{p^2}})]\nonumber \\\label{eq:trans1}
	 &&-{\frac{\partial}{\partial p}}[\dot p\psi-{\frac{p}{3}}(\nabla\cdot V_c)\psi]
	-{\frac{\psi}{\tau_f}}-{\frac{\psi}{\tau_r}}\,, \label{eq:trans2}
\end{eqnarray}
where $\psi$ is the density of CR particles per unit momentum, $q(x,p)$ is the source term, $D_{xx}$ is the spatial diffusion coefficient, $V_c$ is the convection velocity, $D_{pp}$ is the momentum space diffusion coefficient, $\dot p\equiv dp/dt$ describes ionization and Coulomb losses, $\tau_f$ is the time scales for collisions off gas nuclei, and $\tau_r$ is the time scales for radioactive decay.

The scattering of CR particles on randomly moving magnetohydrodynamics (MHD) waves leads to stochastic acceleration, which is described in the transport equation as diffusion in momentum space $D_{pp}$. Considering the scenario where the CRs are reaccelerated by colliding with the interstellar random weak hydrodynamic waves, the relation between the spatial diffusion coefficient $D_{xx}$ and the momentum diffusion
coefficient $D_{pp}$ is expressed as \cite{1994ApJ...431..705S}:

\begin{equation}
	D_{xx}D_{pp}=\frac{4p^2V_a^2}{3\delta(4-\delta)(4-\delta^2)\omega}\,.
\end{equation}

We adopt the numerical {\footnotesize GALPROP} v56\footnote{Current version available at \url{https://galprop.stanford.edu}} framework \cite{Strong:1998pw,Strong:1998fr}.
The reaction network is a series of repeated procedures, starting at solving the propagation equation of the heaviest nuclei $\rm{^{64}_{28}Ni}$, computing all the resulting secondary sources, and then proceeding to the nuclei with $A-1$, where $A$ is the mass number of nuclei. 

The CR injection spectrum is assumed to be a broken power law in rigidity with a low-energy break and a high-energy ($>200$ GV) break, to fit the observed low-energy spectral bumps and the high-energy spectral hardening.
In the {\footnotesize GALPROP} setup, we specify different power-law indices and abundances for injection spectra of nuclei, constrained by the measured AMS-02 data \cite{Consolandi:2016fhd,Aguilar:2017hno,AMS:2020cai,Aguilar:2021tos,AMS:2021brg,AMS:2021lxc} and other experiments \cite{Gorbunov:2018stf,Ahn:2009tb,Adriani:2020wyg}. Table~\ref{tab:abun} illustrates our preliminary fit of these elemental abundances.
\begin{table}
\caption{\label{tab:abun}Source Abundances of CR species, when one fixes the abundance of the proton to $1.06\times10^6$ at 100 GeV$/n$.}
\begin{tabular}{c|c|c|c|c|c|c|c|c|c|c|c}
 Element&C&N&O&Ne&Na&Mg&Al&Si&S&Ca&Fe\\ \hline
 Abund.&3461&253&4548&594&23&868&65&836&93&52&834
\end{tabular}
\end{table}
As visually shown in App.~\ref{app:xsdata}, $\rm{^{20}Ne}$ can produce more cumulative fluorine (half of them decayed from $\rm{^{19}Ne}$) than what $\rm{^{22}Ne}$ can produce. The F/Si ratio decreases when the relative fraction of $\rm{^{22}Ne}$ to $\rm{^{20}Ne}$ at the source grows.
Here we keep the abundance ratio $\rm{(^{22}Ne/^{20}Ne)_{CRS} }= 0.32$ reported by the {\footnotesize GALPROP} team \cite{Boschini:2020jty}, which is obtained from ACE-CRIS Ne isotopic data.
The purely secondary-to-primary ratios are independent of their progenitors’ source spectrum, and we keep the primary abundance and power-law index fixed for simplicity.

\subsection{Cross-section Datasets \label{xs_datasets}}
The setup of the cross section started from a systematic approach tuned to all available data of the most related channels that produce $\rm{^{19}F}$, $\rm{^{10}B}$, or $\rm{^{11}B}$. These cross-section data are taken mostly from {\footnotesize GALPROP}'s core file \texttt{isotope\_cs.dat}, others from  \cite{Zeitlin:2001ye,Zeitlin:2007sm,Zeitlin:2011qg,Villagrasa-Canton:2006exk,FLESCH2001237,Napolitani:2004fw} and the Experimental Nuclear Reaction Data (EXFOR) database\footnote{\url{https://www-nds.iaea.org/exfor/exfor.htm}}.

\begin{table}
\caption{\label{tab:channel}Ranking of the isotopic production cross sections at 10 GeV$/n$.}
\begin{ruledtabular}
\begin{tabular}{cc}
 Channel&contribution [\%]\\ \hline
  $^{20}\text{Ne} \longrightarrow ^{19}\text{Ne}\longrightarrow ^{19}\text{F}$&20.742\\
  $^{24}\text{Mg} \longrightarrow ^{19}\text{F}$&17.716\\
  $^{20}\text{Ne} \longrightarrow ^{19}\text{F}$&17.654\\
  $^{28}\text{Si} \longrightarrow ^{19}\text{F}$&10.493\\
  $^{22}\text{Ne} \longrightarrow ^{19}\text{F}$&6.741\\
   $^{24}\text{Mg} \longrightarrow ^{19}\text{Ne}\longrightarrow ^{19}\text{F}$&3.758\\
  $^{28}\text{Si} \longrightarrow
      ^{19}\text{Ne}\longrightarrow ^{19}\text{F}$&3.341\\
  $^{23}\text{Na} \longrightarrow ^{19}\text{F}$&2.912\\
   $^{27}\text{Al} \longrightarrow ^{19}\text{F}$&2.296\\
   $^{56}\text{Fe} \longrightarrow ^{19}\text{F}$&1.981\\
  $^{22}\text{Ne} \longrightarrow ^{19}\text{Ne}\longrightarrow ^{19}\text{F}$&0.238\\\hline
   $^{16}\text{O} \longrightarrow ^{11}\text{B}(^{11}\text{C})$\footnote{{\footnotesize GALPROP}'s fragmentation routine regards ghost nucleon $^{11}\text{C}$ produced from $\rm{^{16}O}$ and $\rm{^{14}N}$ as $^{11}\text{B}$, which is the same for ghost nuclei $^{10}\text{C}$. See also in App.~\ref{app:xsdata}.}&20.491\\
   $^{12}\text{C} \longrightarrow ^{11}\text{B}$&20.346\\  
    $^{12}\text{C} \longrightarrow ^{11}\text{C} \longrightarrow ^{11}\text{B}$&18.059\\  
    $^{12}\text{C} \longrightarrow ^{10}\text{B}$&8.225\\  
  $^{16}\text{O} \longrightarrow ^{10}\text{B}(^{10}\text{C})$&8.121\\
  $^{14}\text{N}\longrightarrow ^{11}\text{B}(^{11}\text{C})$&2.601\\
 $^{15}\text{N}\longrightarrow ^{11}\text{B}$\footnote{Most $^{15}\text{N}$ flux comes from the production of  $^{16}\text{O}$.}&2.177

\end{tabular}
\end{ruledtabular}
\end{table}
The nuclear reaction network is built by {\footnotesize GALPROP} using the Nuclear Data Sheets, and we have isolated the contributions for ranking the important reactions. Projectile $X$ on interstellar medium (ISM) target (mostly H or He) producing a fragment $F$ can be written as $X + \{p, \alpha\} \longrightarrow F$.
The dominant channels in multi-step reactions can be estimated by forming a product of the relative abundance of CR species and the associated production cross sections.
 Each channel's contribution changes differently with energy, and we estimate them at 10 GeV$/n$, where most cross sections remain energy-independent.
The isotropic $^{19}\text{Ne}$, $^{10}\text{C}$, and $^{11}\text{C}$ are generally considered as ghost nuclei, since these short-lived intermediate nuclei decay quickly into $^{19}\text{F}$, $^{10}\text{B}$ and $^{11}\text{B}$ before interacting with interstellar gas. However, the proportion of the ghost nuclei contributing to the cumulative cross section for a particular reaction is an important quantity, and we should not ignore them.

Following this routine, we have illustrated the possible contributions of every channel and their contribution rates in Table~\ref{tab:channel}.
As listed in the table, $^{12}\text{C}$, $^{16}\text{O}$, and $^{14}\text{N}$ are the main projectiles contributing secondary B, as they have rich abundances, and their production cross sections are relatively large.
For almost the same reason, $^{20}\text{Ne}$, $^{24}\text{Me}$, and $^{28}\text{Si}$ are the main projectiles contributing secondary F.
We choose to ignore three channels: $^{27}\text{Al} \longrightarrow ^{19}\text{F}$, $^{56}\text{Fe} \longrightarrow ^{19}\text{F}$, and $^{22}\text{Ne} \longrightarrow ^{19}\text{Ne}\longrightarrow ^{19}\text{F}$, since their uncertainties hardly affect the F flux.

We adopt several different parametrizations for the cross sections to discuss the systematic uncertainties.  
These original parametrizations are generated by different assumptions and approaches to fit similar datasets.
For instance, the {\footnotesize GALPROP} parametrization is normalized to data for specific reaction whenever available, whereas other parametrizations are usually based on semi-empirical formulae designed to give a self-consistent fit over different reactions. The former is expected to fit the data better. However, it sometimes shows step-like energy dependencies that may not be physical, while the other parametrizations do not. In this work, we used four kinds of parametrizations provided by {\footnotesize GALPROP} code---WE93, TS00, GAL12, and GAL22 \cite{Genolini:2018ekk,Villagrasa-Canton:2006exk}:

WE93 is one of Webber's parametrizations based on the 1993 version of \texttt{WNEWTR.FOR} code. The parametrizations developed by W.R.Webber and co-workers \cite{Webber:1990kc,Webber:1998ex,Webber_2003} are based on certain observations of nuclear fragmentation.
The formulae have three terms:
(i) the charge-changing term, with an exponential
dependence on the charge difference between the
projectile and the fragment;
(ii) the isotopic term, where the width of the individual mass for a given fragment charge follows the Gaussian distribution;
(iii) the energy dependence term, which describes the overall energy dependence of the particular cross section.
 The phenomenological approximation of cross-section formulae is normalized to the data at 600 MeV$/n$, then the energy dependencies are applied to fragments of similar charge. 

TS00 is a semi-empirical parametrization
developed by Tsao and Silberberg \cite{Silberberg:1998lxa} in their 2000 version of \texttt{YIELDX 011000.FOR} code, based on the data from the Transport collaboration.
The formulae rely on multiple regularities of nuclei,
including mass differences between the projectile and the fragment,
the ratio of the number of neutrons and protons,
the level structure of the residual nuclei, the factor for the pairing of protons and neutrons, and so on.

GAL12 and GAL22 are developed in {\footnotesize GALPROP} code as the default parametrizations 
based on a careful inspection of the quality and systematics of various datasets and semi-empirical formulae.
The isotopic cross sections are calculated using the fits given by \cite{Strong:2001fu,Moskalenko:2001qm,Moskalenko:2003kp} to major beryllium and boron production cross sections, usually a direct fit to the data for each particular reaction.
For less important cross sections, the phenomenological approximations by WE93 (option 12 in {\footnotesize GALPROP}) or TS00 (option 22 in {\footnotesize GALPROP}) are
used instead, normalized to the data if available.

The results in Table~\ref{tab:channel} are based on the GAL12 parametrization.
Changing the parametrization would slightly change the fraction of contributions, especially for GAL22 and TS00 which assume more productions for heavier nuclei ($A>36$) into F. Unfortunately, we cannot analyze the data-driven uncertainties for these heavier projectiles, as the corresponding data are scarce or none. As a rough estimation, we do the fit below by using four parametrizations, so that different predictions of heavier nuclei into F can be considered. We refer the reader to \cite{Genolini:2018ekk} for a systematic analysis of the channels contributing to different nuclei. Their analysis of isotropic boron (TABLE~XI in \cite{Genolini:2018ekk}) is quite similar to ours.

Note that all the four parametrizations provided by the {\footnotesize GALPROP} code have been normalized to more complete datasets than those in the earlier works. Therefore, they may be different from the original parametrizations in \texttt{WNEWTR.FOR} code and \texttt{YIELDX 011000.FOR} code, or these provided in {\footnotesize USINE} code\footnote{\url{https://lpsc.in2p3.fr/usine}}.

\section{Fitting RESULTS\label{sec:result}}
\subsection{Preliminary Fitting for Each parametrizations \label{subsec:prefit}}
In Appendix~\ref{app:xsdata}, all collected data above 100 MeV$/n$ are plotted together with the parametrizations taken from {\footnotesize GALPROP} code. 
To determine the cross-section uncertainties, we perform a
data-driven renormalization on the benchmark parametrizations, where all cross sections are assumed to be asymptotically energy-independent above several GeVs. Models of CR propagation rely seriously on these extrapolations, while different parametrizations may predict different extrapolations.

Our approach follows the data-driven procedure of earlier
studies \cite{Moskalenko:2013heu,Tomassetti:2015nha,Tomassetti:2017hbe}. We re-normalize those formulae to available data points by applying overall re-scalings\footnote{The energy scale is ignored since it modifies the cross section blow the energy region where we are interested. This approach possibly underestimates the uncertainty.}.
We demonstrate the result as Gaussian distributed parameters with central value $\mu$ and dispersion $\sigma$. 
Table~\ref{tab:gaussian-fit} lists all the best-fit values of Gaussian distributed parameters for each parametrization.
Cross-section channels for GAL12 and WE93 show the same Gaussian distributed parameters when their fragment is F, as they share the same formulae when data are less collected. GAL22 and TS00 have a similar relationship.
Other channels with more data for GAL12 and GAL22 parametrizations are almost the same, obeying a direct fit to the data when their fragment is B.
Since we have added new data for renormalization, the central values $\mu$ of some updated reactions are within a few percent below or above the old ones,
such as $^{28}\text{Si} \longrightarrow ^{19}\text{F}$, $^{24}\text{Mg} \longrightarrow ^{19}\text{Ne}\longrightarrow ^{19}\text{F}$, and $^{28}\text{Si} \longrightarrow ^{19}\text{Ne}\longrightarrow ^{19}\text{F}$.

\begin{table*}
\caption{\label{tab:gaussian-fit}The best-fit values of Gaussian distributed parameters for every parametrization}
\begin{ruledtabular}
\begin{tabular}{ccccc}
 Channel&GAL12&GAL22&TS00&WE93\\ \hline
  &$\mu - \sigma$&$\mu - \sigma$&$\mu - \sigma$&$\mu - \sigma$\\
  $^{20}\text{Ne} \longrightarrow ^{19}\text{Ne}\longrightarrow ^{19}\text{F}$&1.000 - 0.035&1.000 - 0.035&1.000 - 0.035&1.000 - 0.035\\
   $^{24}\text{Mg} \longrightarrow ^{19}\text{F}$&0.998 - 0.057&0.985 - 0.056&0.985 - 0.056&0.998 - 0.057\\
 $^{20}\text{Ne} \longrightarrow ^{19}\text{F}$&1.005 - 0.027&1.010 - 0.027&1.010 - 0.027&1.005 - 0.027\\
    $^{28}\text{Si} \longrightarrow ^{19}\text{F}$&1.073 - 0.054&1.072 - 0.054&1.072 - 0.054&1.073 - 0.054\\
   $^{22}\text{Ne} \longrightarrow ^{19}\text{F}$&1.000 - 0.039&1.000 - 0.039&1.000 - 0.039&1.000 - 0.039\\  
    $^{24}\text{Mg} \longrightarrow ^{19}\text{Ne}\longrightarrow ^{19}\text{F}$&1.042 - 0.123&1.044 - 0.123&1.044 - 0.123&1.042 - 0.123\\
 $^{28}\text{Si} \longrightarrow
 ^{19}\text{Ne}\longrightarrow ^{19}\text{F}$&1.057 - 0.085&1.070 - 0.087&1.070 - 0.087&1.057 - 0.085\\   
  $^{23}\text{Na} \longrightarrow ^{19}\text{F}$&1.000 - 0.200&0.998 - 0.200&0.998 - 0.200&1.000 - 0.200\\ \hline
   $^{16}\text{O} \longrightarrow ^{11}\text{B}(^{11}\text{C})$&0.964 - 0.026&0.964 - 0.026&1.023 - 0.027&1.021 - 0.027\\  
   $^{12}\text{C} \longrightarrow ^{11}\text{B}$&0.956 - 0.019&0.956 - 0.019&0.998 - 0.019&0.994 - 0.019\\   
 $^{12}\text{C} \longrightarrow ^{11}\text{C} \longrightarrow ^{11}\text{B}$&1.012 - 0.008&1.012 - 0.008&1.080 - 0.008&1.021 - 0.008\\ 
    $^{12}\text{C} \longrightarrow ^{10}\text{B}$&0.996 - 0.033&0.996 - 0.033&1.000 - 0.033&0.992 - 0.033\\  
  $^{16}\text{O} \longrightarrow ^{10}\text{B}(^{10}\text{C})$&1.036 - 0.030&1.036 - 0.030&0.973 - 0.028&0.970 - 0.028\\  
     $^{14}\text{N}\longrightarrow ^{11}\text{B}(^{11}\text{C})$&1.025 - 0.028&1.025 - 0.028&0.958 - 0.026&0.944 - 0.026\\
     $^{15}\text{N}\longrightarrow ^{11}\text{B}$&1.000 - 0.045&1.000 - 0.045&1.001 - 0.044&1.000 - 0.045\\
     
\end{tabular}
\end{ruledtabular}
\end{table*}
\begin{figure}[htbp]
\includegraphics[width=0.5\textwidth,trim=150 50 0 0,clip]{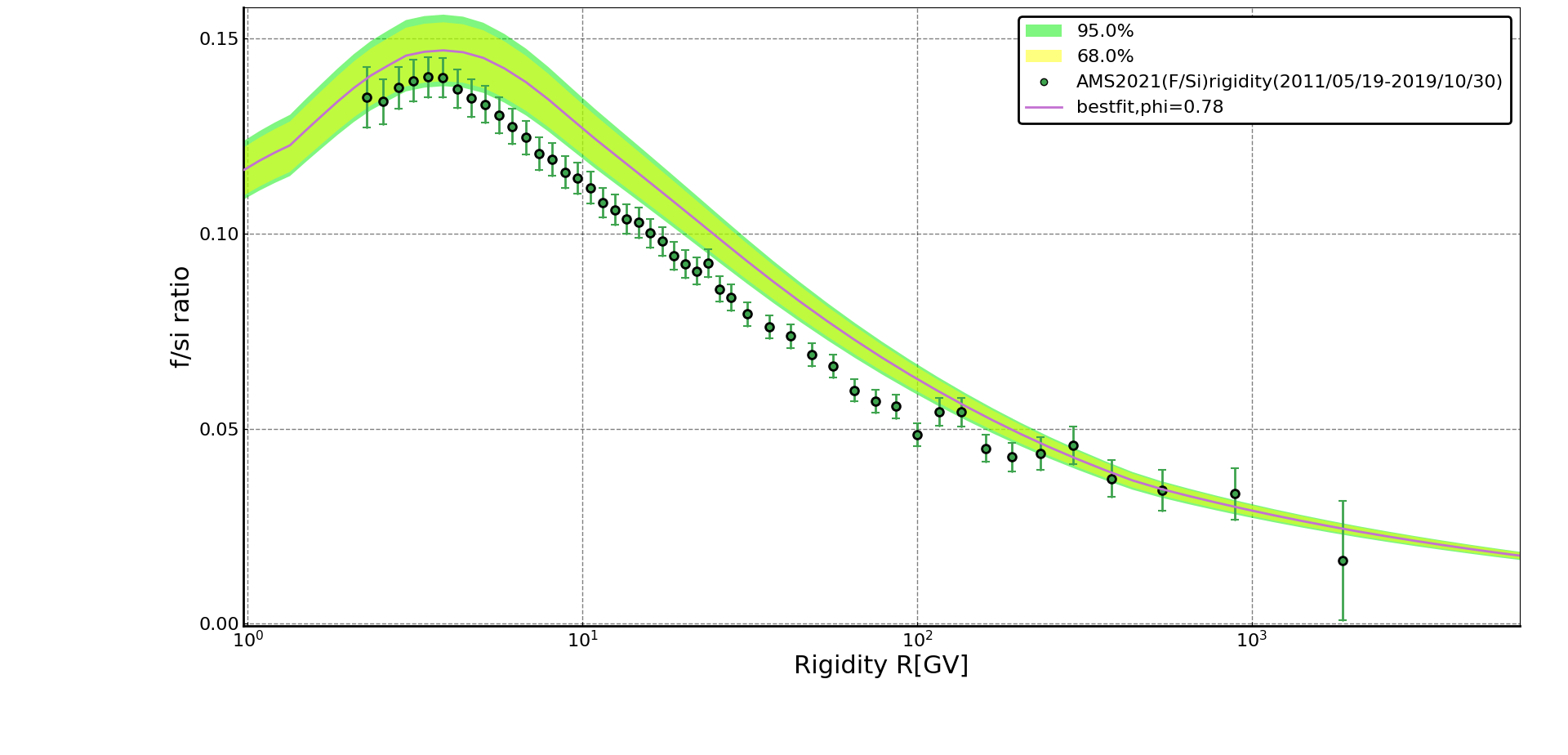}\\
\caption{\label{fig:sigmaf}F/Si ratio predicted by the B/C best-fit parameters (purple line) and the confidence interval (yellow and green band) caused by cross-section uncertainties,
compared with the experimental data from AMS-02 \cite{AMS:2021tnd}.}
\end{figure}
As shown in Fig.~\ref{fig:sigmaf}, using the diffusion
coefficient inferred from the B/C ratio reported by AMS-02 \cite{Aguilar:2021tos}, the prediction of the F/Si ratio (purple line) shows a significant excess below 100 GV. Although we draw the 1$\sigma$ and 2$\sigma$ error bars according to the data restriction listed in Table~\ref{tab:gaussian-fit} for all the channels producing $\rm{^{19}F}$, the possible uncertainties do not seem to account for the overestimation.
Here we only illustrate the result of the GAL12 parametrization,
while the overestimation commonly emerges in other parametrizations.

\subsection{Combined fitting of nucleon ratios and cross sections\label{sec:comb}}
Considering the cross-section uncertainties, we would like to test if B/C and F/Si can be interpreted simultaneously. For the goodness-of-fit of the model to the data, we use the least-$\chi^2$ method. The $\chi^2$ statistic is expressed as
\begin{equation}
\label{eq:chi}
  \chi^2=\sum_{q=1}^{2}\mathcal{D}_q+\chi^2_{\rm{cs}},
\end{equation}
\begin{equation}
\label{eq:data}
  \mathcal{D}=\sum_{i=1}^{\rm{bin}}(\frac{y_i^{\rm{data}}-y_i^{\rm{model}}}{\sigma_i^{\rm{data}}})^2,
\end{equation}
\begin{equation}
\label{eq:chi_cs}
  \chi^2_{\rm{cs}}=\sum_{i=1}^{n_{\rm{cs}}}(\frac{\mu_i-C_i}{\sigma_i})^2.
\end{equation}
In the above equations, the $q$ runs over B/C and F/Si, and each energy bin is calculated separately to get the quadratic distance between the data and the model.
The constraint from the cross-section data contributes an additional term $\chi^2_{\rm{cs}}$,
where $\mu_i$ and $\sigma_i$ are the central value and dispersion of the specific channel $i$, and $C_i$ is the tested value of renormalization in the fit.
Note that the CR data contributes most of the degrees of freedom (d.o.f.) and has better accuracy than the cross-section data. Thus, it is possible that the best-fit result has an acceptable $\chi^2/n_{\rm{d.o.f.}}$ but a $\chi^2_{\rm{cs}}/n_{\rm{cs}}$ significantly larger than $1$, where $n_{\rm{d.o.f.}}$ is the total d.o.f. and $n_{\rm{cs}}$ is the d.o.f. of the cross-section data.

MCMC methods are widely used in Bayesian inference and are powerful to sample the multi-dimensional parameter space for CR propagation models \cite{Masi:2016uby,Putze:2010zn,Yuan:2017ozr,Johannesson:2016rlh}.
We perform a combined fitting of nucleon ratios together with cross-section uncertainties by using {\footnotesize CosmoMC}\footnote{\url{https://cosmologist.info/cosmomc}}.
In a previous paper \cite{PhysRevD.104.123001} we have introduced the basic settings for {\footnotesize GALPROP} and {\footnotesize CosmoMC}.
For the resolution of the {\footnotesize GALPROP} calculation, we set a 2D spatial grid of $dr=1$~kpc and $dz=0.5$~kpc, and an energy grid of Ekin\_factor = 1.2, considering both accuracy and speed. The size of the initial time step (start\_timestep) is set to be 1.0e8, which is smaller than the default. We have checked that it does not affect the results. Other parameters are kept as the defaults of {\footnotesize GALPROP}~v56.

The group of free parameters in the fitting procedure is 
\begin{eqnarray*}
    \bm{\theta}=\{D_0,\delta,V_a,\eta,
    xs_{20ne-19ne},xs_{24mg-f},xs_{20ne-f},...\},
\end{eqnarray*}
where the $D_0$, $\delta$, $V_a$, and $\eta$ are the propagation parameters; the rest parameters are the renormalization factors for the cross sections of specific channels. As mentioned in Sec.~\ref{subsec:proper}, we have preliminary adjusted the individual isotopic source abundance and slope index to match the corresponding primary fluxes from AMS-02. There is a strong degeneracy between the diffuse coefficient and the halo height $L$. We fix $L$ to be $6$ kpc, which is a conservative value consistent with various studies \cite{Weinrich:2020ftb,Putze:2010zn,Trotta:2010mx,Evoli:2019iih} without loss of generality. 

We adopt the force-field approximation for the solar modulation \cite{Gleeson:1968zza}, where the strength is described by the solar modulation potential $\phi$. Since the measurements used were taken during a similar period (B/C taken during May 2011-May 2018 \cite{Aguilar:2021tos} and F/Si taken during May 2011-Oct 2019 \cite{AMS:2021tnd}), we assume that they share the same $\phi$. 
We find that the impact of solar modulation on the secondary-to-primary ratios is little and fix $\phi=0.76$ GV for both the cases, which is obtained from the fits considering the un-modulated Voyager data (see Sec.~\ref{sec:effective}).

\begin{figure}[htbp]
\includegraphics[width=0.5\textwidth]{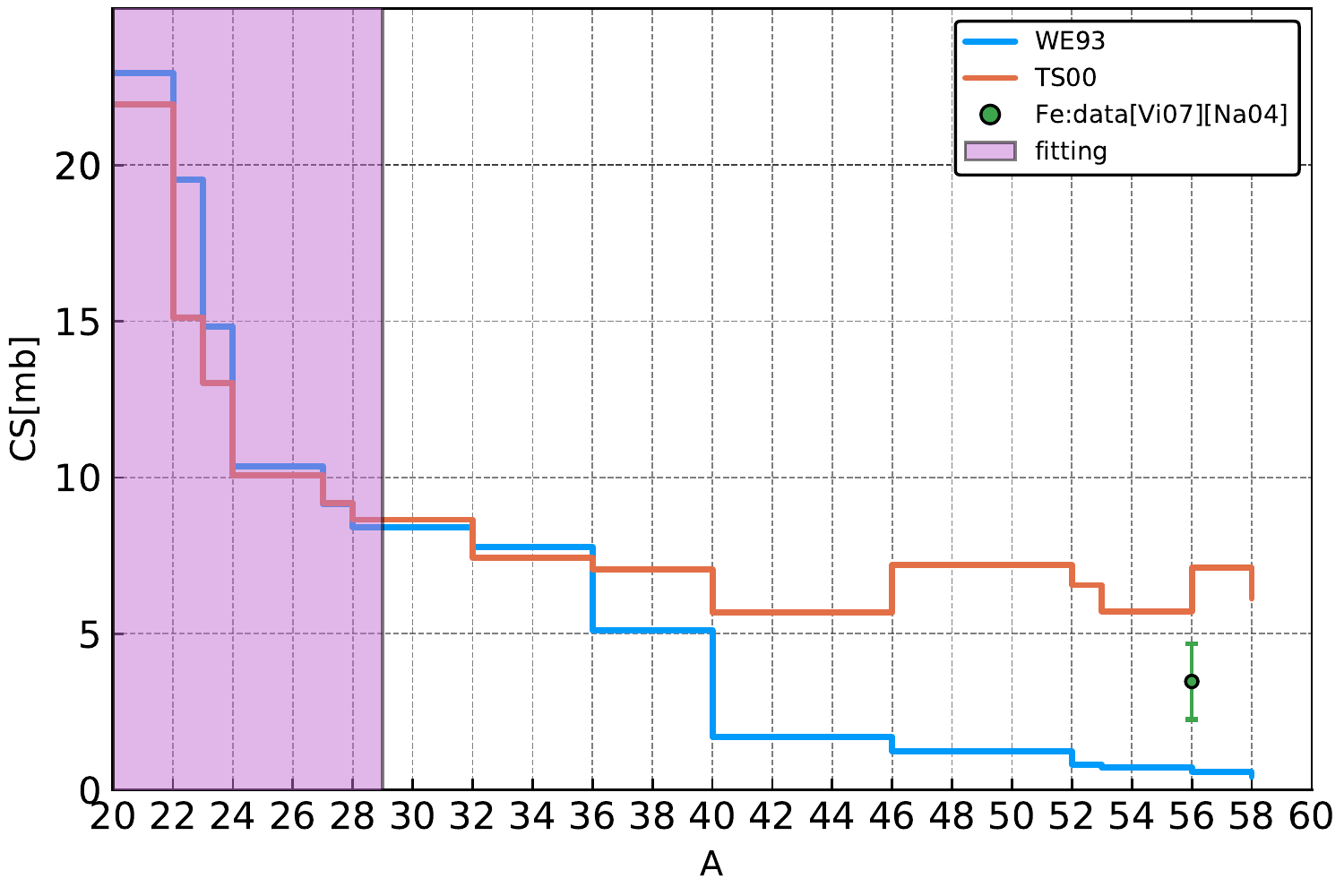}\\
 \includegraphics[width=0.5\textwidth]{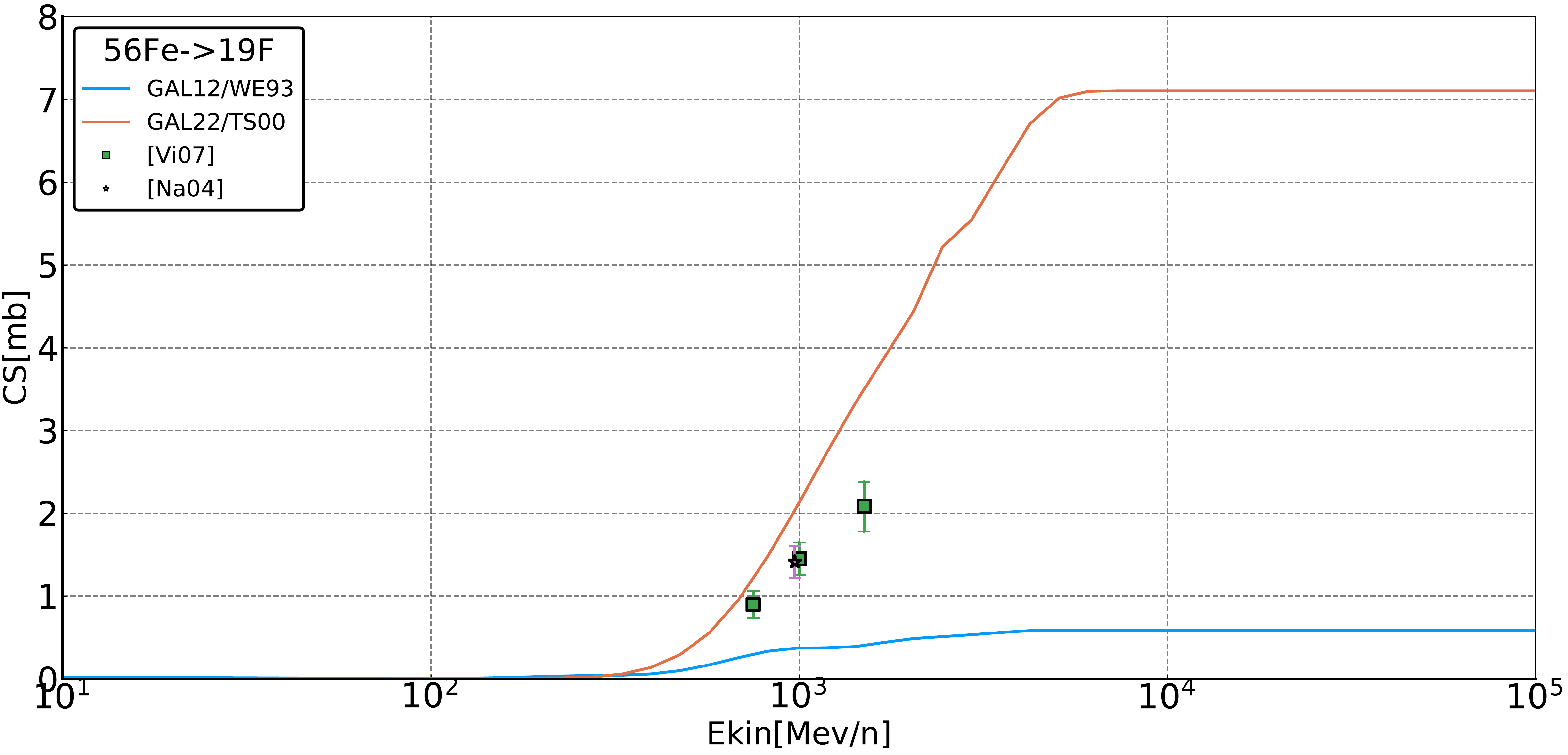}\\
\caption{\label{fig:tswe} 
Top: Mass distribution of the projectile nuclei in the spallation reaction $X + p \longrightarrow \rm{^{19}F}$ at 10 GeV$/n$, compared between parametrizations (lines) and estimated data (green point). The purple shaded area is where we re-normalize the formulae for combined fitting.
Bottom: Comparison of parametrizations (lines) and nuclear data (symbols) for the fragmentation cross sections of $\rm{^{56}Fe}$ into $\rm{^{19}F}$. Data are from \cite{Villagrasa-Canton:2006exk,Napolitani:2004fw}.}
\end{figure}

The fitting result is shown in Table~\ref{tab:combined}. The goodness of fit for the GAL12 and WE93 cases are quite good with $\chi^2/n_{d.o.f.}\approx1$, while the GAL22 and TS00 cases fit poor to the data, which are disfavored by confidence levels of more than $99.9\%$. The significant difference is mainly attributed to the different dependency of the cross section on the projectile mass number between WE93 and TS00, as shown in the top panel of Fig.~\ref{fig:tswe}. The TS00 case (red line) prefers to predict more secondary F from heavier nuclei with $A>36$, which will make the problem of F anomaly more serious. 
 
\begin{figure}[htbp]
\includegraphics[width=0.5\textwidth,trim=150 50 0 0,clip]{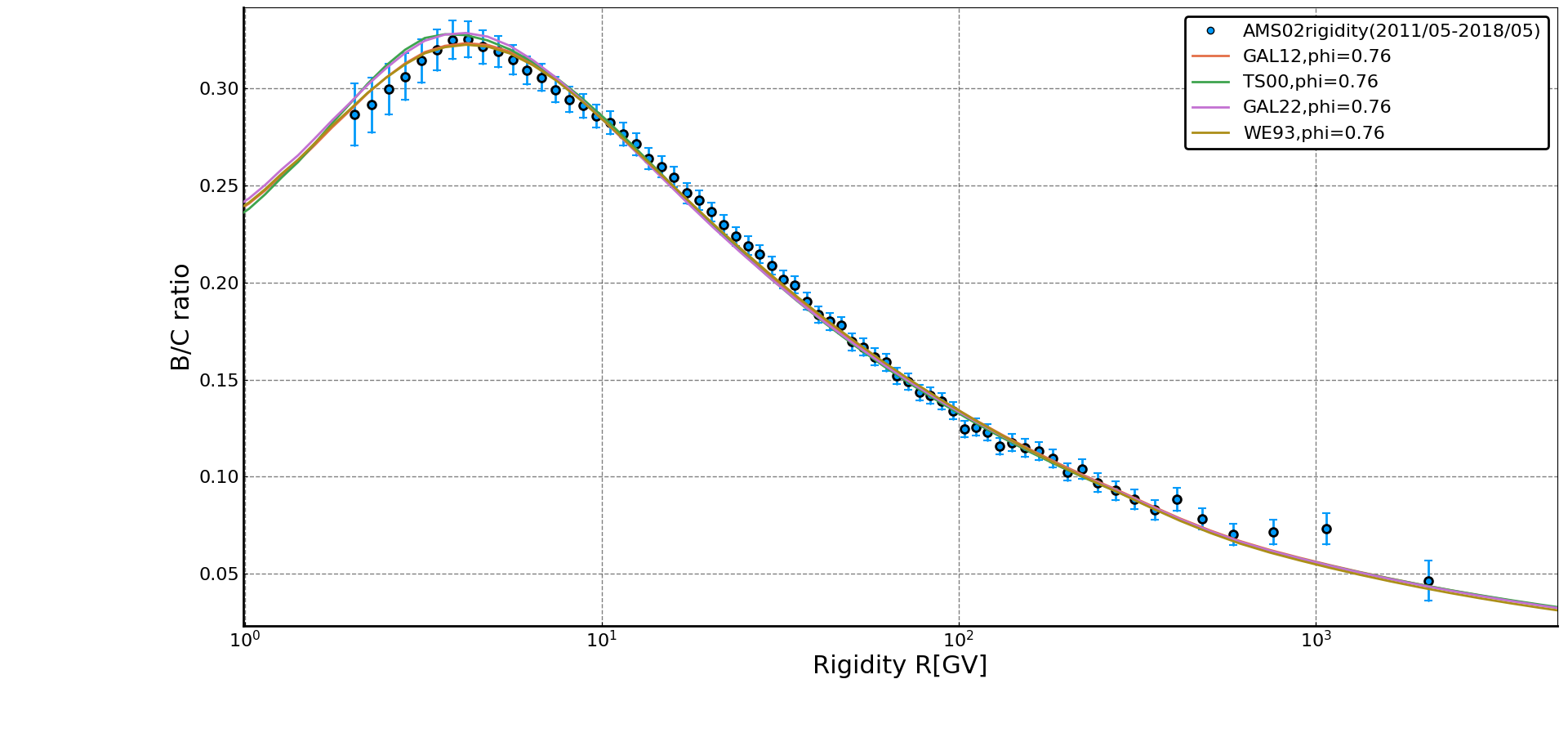}\\
\includegraphics[width=0.5\textwidth,trim=150 50 10 0,clip]{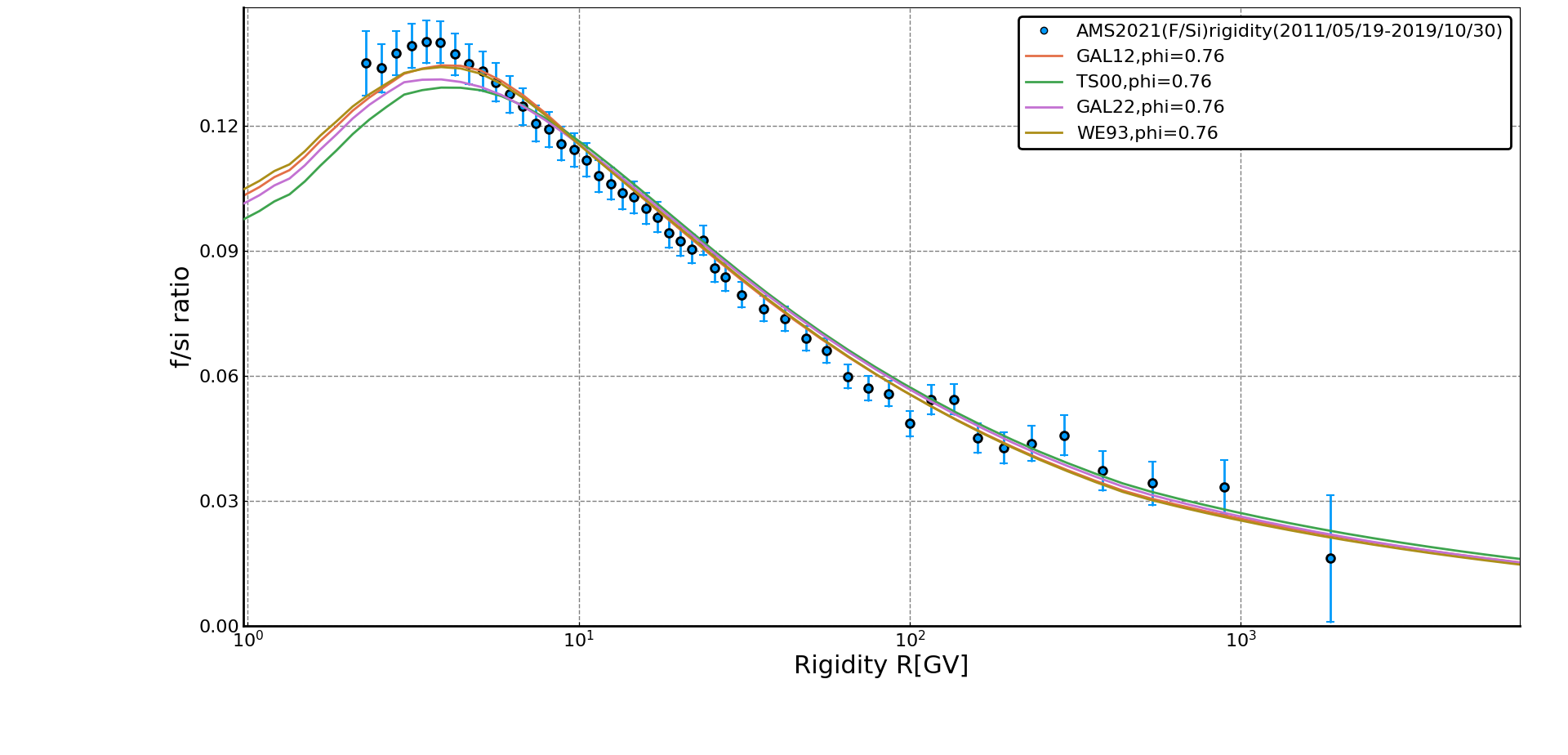}\\
\caption{\label{fig:bc-fsi} 
Comparison of data \cite{Aguilar:2021tos,AMS:2021tnd} with calculations of the best-fit parameters based on the four parametrizations (red line for GAL12, green line for TS00, purple line for GAL22, and brown line for WE93).
Top: Result of B/C ratio for different parametrizations. Bottom: Result of F/Si ratio for different parametrizations.}
\end{figure}

The best-fit results of B/C and F/Si for all the parametrization are shown in Fig.~\ref{fig:bc-fsi}. Visually all these four parametrizations fit the CRs data well, while the predicted F/Si below 10 GV is significantly less than the data points of AMS-02, especially for TS00 and GAL22. The cross-section uncertainties may not account for the flux excess at low energies as also noticed by \cite{Boschini:2021ekl,2022arXiv220801337F}. The bump-like energy dependence assumed by the cross-section formulae are mostly below 1 GeV$/n$, corresponding to the $\rm{^{19}F}$ rigidity below 3 GV, and the formulae remain nearly unchanged as the energy increases. If the cross-section renormalizations are adjusted to interpret the low-energy F/Si, the high-energy overestimation will be more serious. Primary F is proposed to interpret the low-energy excess \cite{Boschini:2021ekl}, while \cite{2022arXiv220801337F} attributed that excess to the systematic uncertainties of the AMS-02 data, which could make the low-energy bin correlations stronger.

\begin{table}
\caption{\label{tab:combined}Best-fit values of all parameters in the combined fitting}
\begin{ruledtabular}
\begin{tabular}{ccccc}
 Parameter&GAL12&GAL22&TS00&WE93\\ \hline
 $D_0(10^{28}~\rm{cm^2s^{-1}})$&5.879&6.618&6.984&5.691\\
 $\delta$&0.421&0.413&0.400&0.433\\
 $V_a$ (km/s)&25.195&25.789&28.694&22.618\\
 $\eta$&-0.164&0.103&0.028&-0.308\\ \hline
 $xs_{20ne-19ne}$&0.920&0.876&0.904&0.937\\
 $xs_{24mg-f}$&0.810&0.749&0.768&0.837\\
 $xs_{20ne-f}$&0.985&0.966&0.967&0.961\\
 $xs_{28si-f}$&0.972&0.914&0.895&0.990\\
 $xs_{22ne-f}$&0.934&0.915&0.998&0.969\\
 $xs_{24mg-19ne}$&0.824&0.842&0.791&0.764\\
 $xs_{28si-19ne}$&1.030&0.925&0.980&0.995\\
 $xs_{23na-f}$&0.724&0.872&0.683&0.794\\
 $xs_{16o-11bc}$&0.998&1.037&1.113&1.053\\
 $xs_{12c-11b}$&0.973&0.979&1.023&1.029\\
 $xs_{12c-11c}$&1.018&1.016&1.097&1.026\\
 $xs_{12c-10b}$&1.018&1.006&1.018&1.027\\
 $xs_{16o-10bc}$&1.040&1.041&0.985&0.977\\
 $xs_{14n-11bc}$&1.042&1.046&0.965&0.954\\
 $xs_{15n-11b}$&0.989&0.995&1.039&1.010\\
 $\chi^2_{\rm{min}}/n_{\rm{d.o.f.}}$&112.06/111&175.83/111&193.32/111&108.00/111\\
  $\chi^2_{\rm{cs}}/n_{\rm{cs}}$&31.95/15&61.96/15&61.32/15&29.95/15\\
\end{tabular}
\end{ruledtabular}
\end{table}

\begin{figure}[htbp]
\includegraphics[width=0.5\textwidth]{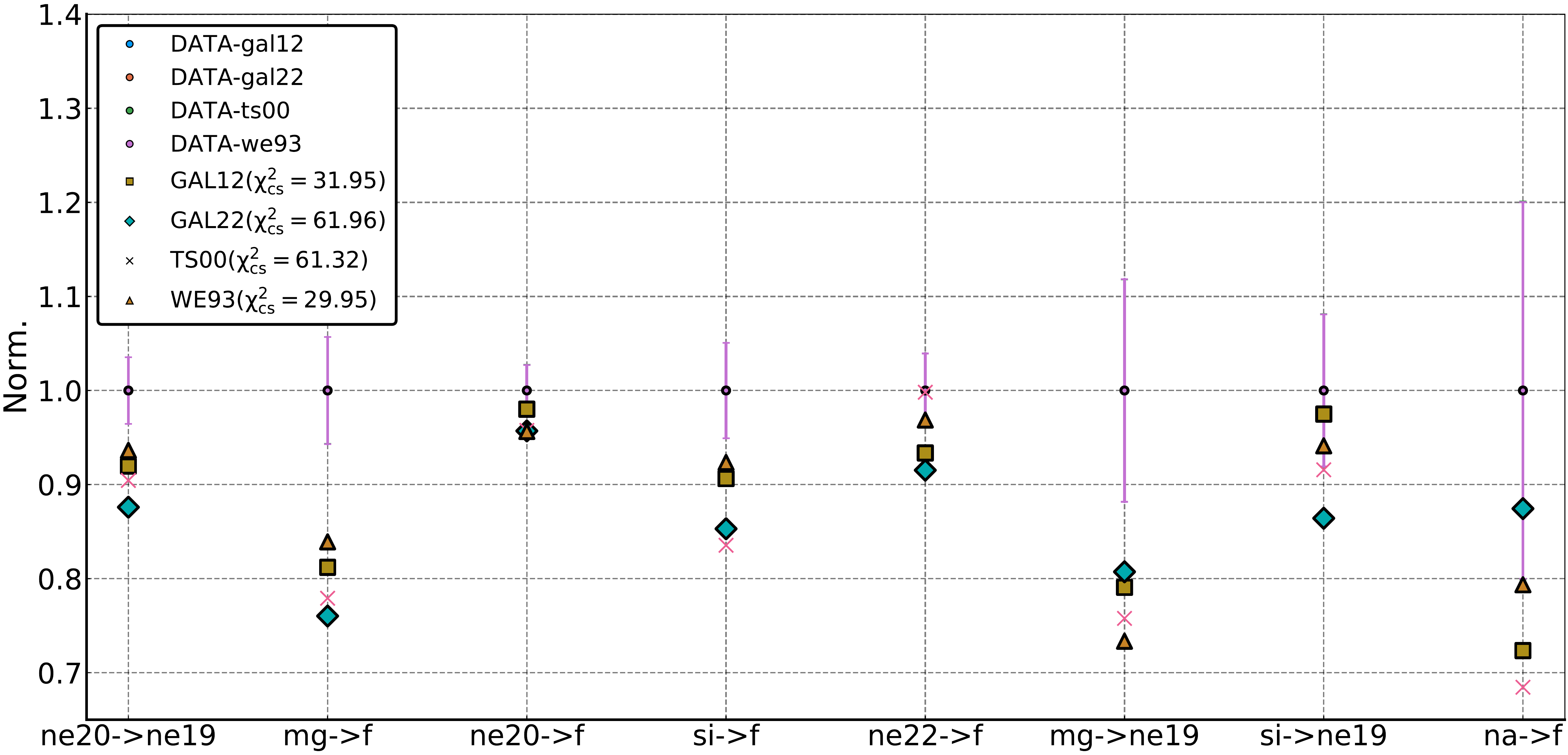}\\
\includegraphics[width=0.5\textwidth]{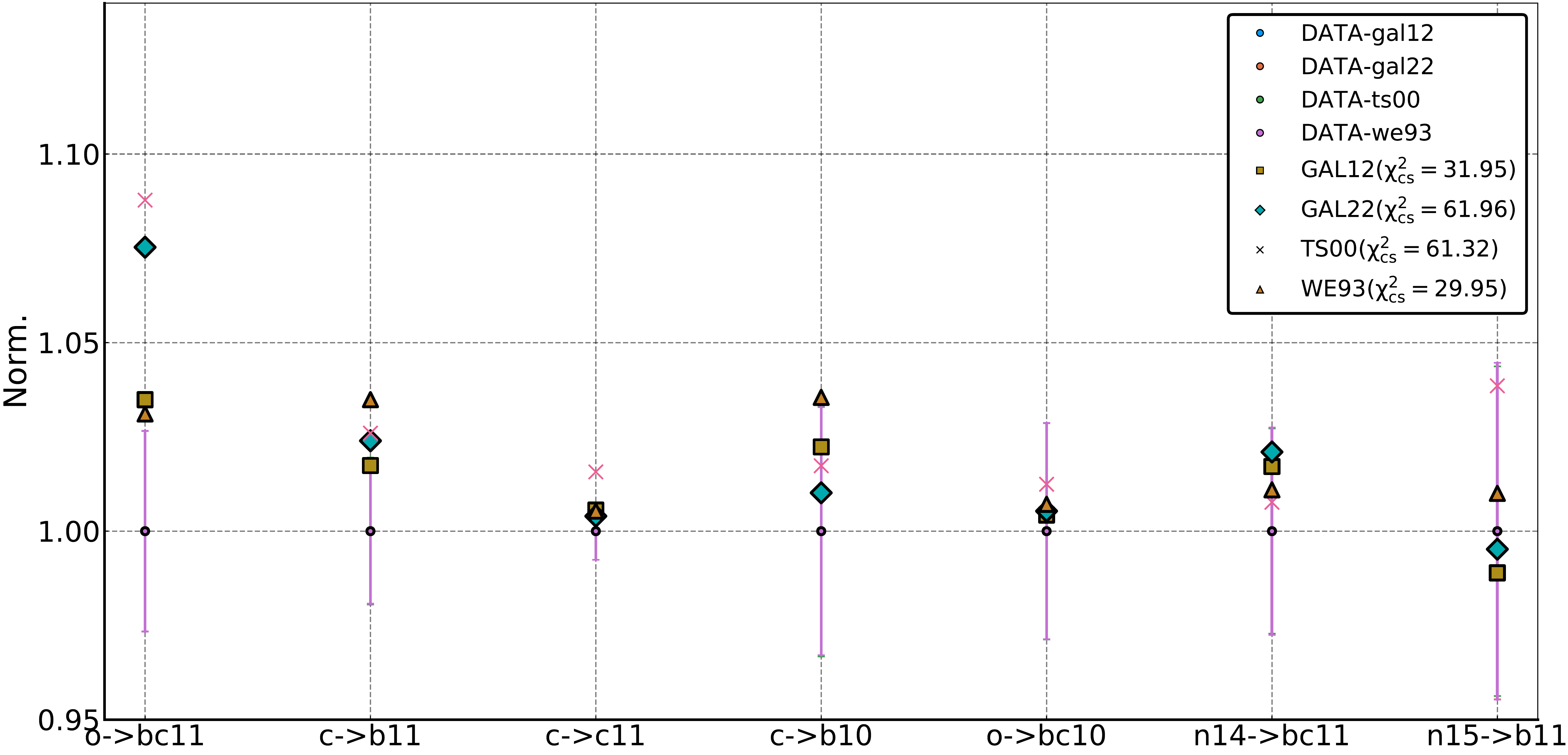}\\
\caption{Comparison of cross-section data expectation from Table~\ref{tab:gaussian-fit} with best-fit renormalization parameters for different parametrizations, marked with different symbols: GAL12 (square), GAL22 (diamond), TS00 (cross), WE93 (triangle). Top: Cross section channels of different projectiles into F.
Bottom: Cross section channels of different projectiles into B.}
\label{fig:xsresult}
\end{figure}

We notice that although the total goodness of fit for the GAL12 and WE93 cases are good, the goodness of fit to the cross-section data is poor with $\chi^2_{\rm{cs}}/n_{\rm{cs}}\gtrsim2$. In Fig.~\ref{fig:xsresult}, we have listed the best-fit results of cross-section renormalization parameters (points without error bars) compared with the values constrained by the experiments (points with error bars). Note that we have re-normalized all these points and errors to the Gaussian means of each channel to show the offsets.
These channels are arranged by their contribution from left to right according to Table~\ref{tab:channel}. Most of the best-fit values deviate from their experimental values by more than $1-2\sigma$. Moreover, all the renormalization parameters for the channels producing F undershoot the experimental values, while the most important channels producing B (to be read from left to right) overshoot the experimental values. The above features suggest that the problem of F anomaly cannot be interpreted by the random errors of the cross sections. The problem also cannot be solved by adopting different existing cross-section models.

As illustrated in Appendix~\ref{app:xsdata}, the energy dependency of the cross section can largely influence the fragment production at higher energies. Hypothetical parametrizations with rapidly decreasing cross sections from 1 to 10 GeV$/n$ for the channels producing F could systematically suppress the F production above 10 GeV$/n$, which may help to interpret the F anomaly. However, most of the cross-section measurements are lower than 1 GeV$/n$. Multi-GeV data are essential to test the possibility of this kind of hypothesis.

An alternative solution is to adjust the dependency of the cross section on the projectile mass number for heavier nuclei.
The different dependencies of WE93 and TS00 shown in the top panel of Fig.~\ref{fig:tswe} have made a significant difference in the goodness of fit. Constructing a parametrization that assumes smaller cross-section production for those unknown reactions can suppress the F flux. However, this hypothesis may be disfavored by some recent observations. 
In the bottom panel of Fig.~\ref{fig:tswe}, the data restriction for $\rm{^{56}Fe}$ into $\rm{^{19}F}$ \cite{Villagrasa-Canton:2006exk,Napolitani:2004fw} lies between the prediction of TS00 (red line) and WE93 (blue line).
By comparing the measured data in \cite{Villagrasa-Canton:2006exk} with the formulae of parametrizations, we find that the overestimation of TS00 and underestimation of WE93 are common for $\rm{^{56}Fe}$ projectile into most lighter fragments. It means that the WE93 parametrization may already underestimate the F flux contributed by the heavy progenitors, and a more reasonable parametrization may make the problem of F anomaly more serious. 

\section{Effective Propagation Distance and Spatially Dependent Diffusion\label{sec:effective}}

\begin{figure}[htbp]
\includegraphics[width=0.5\textwidth]{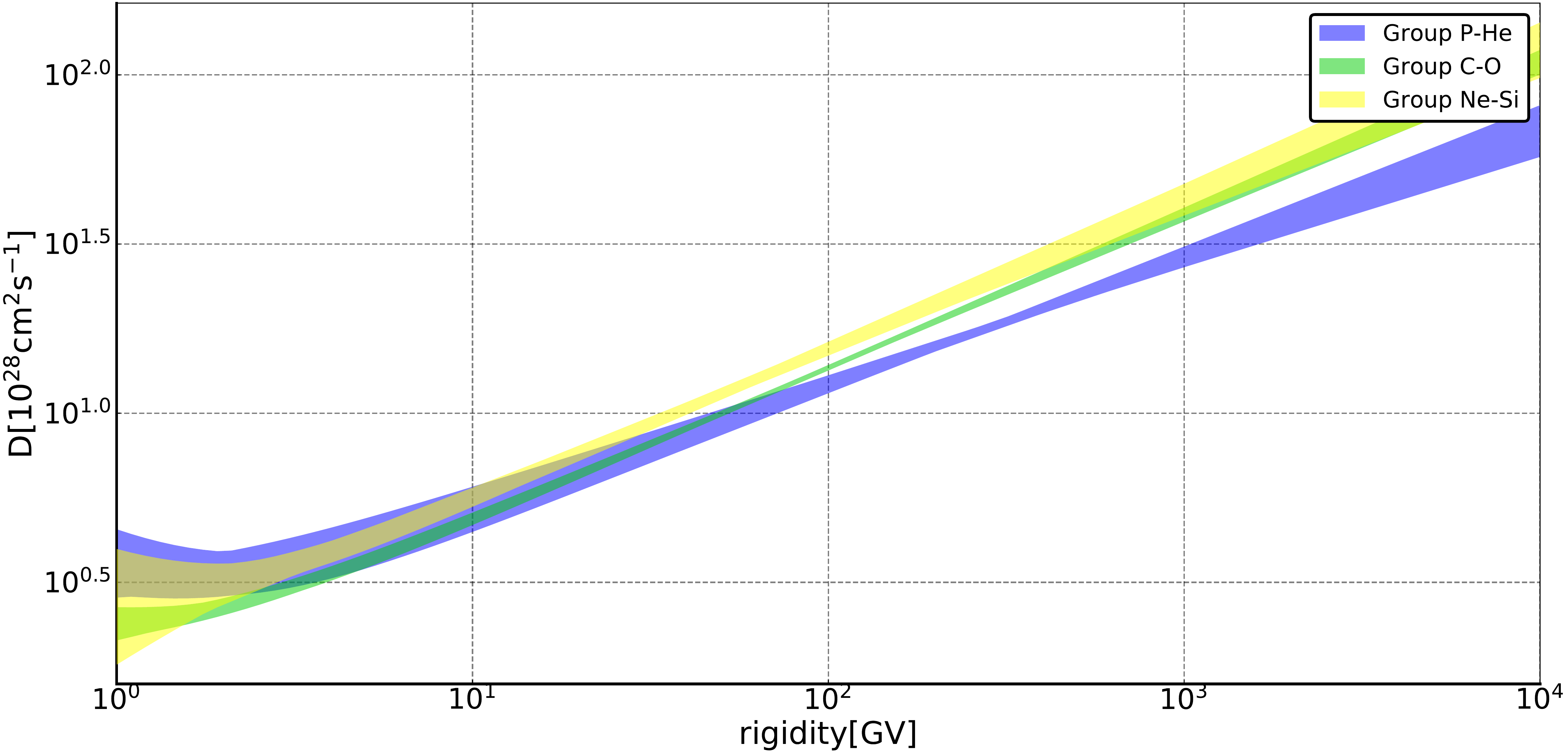}\\
\includegraphics[width=0.5\textwidth]{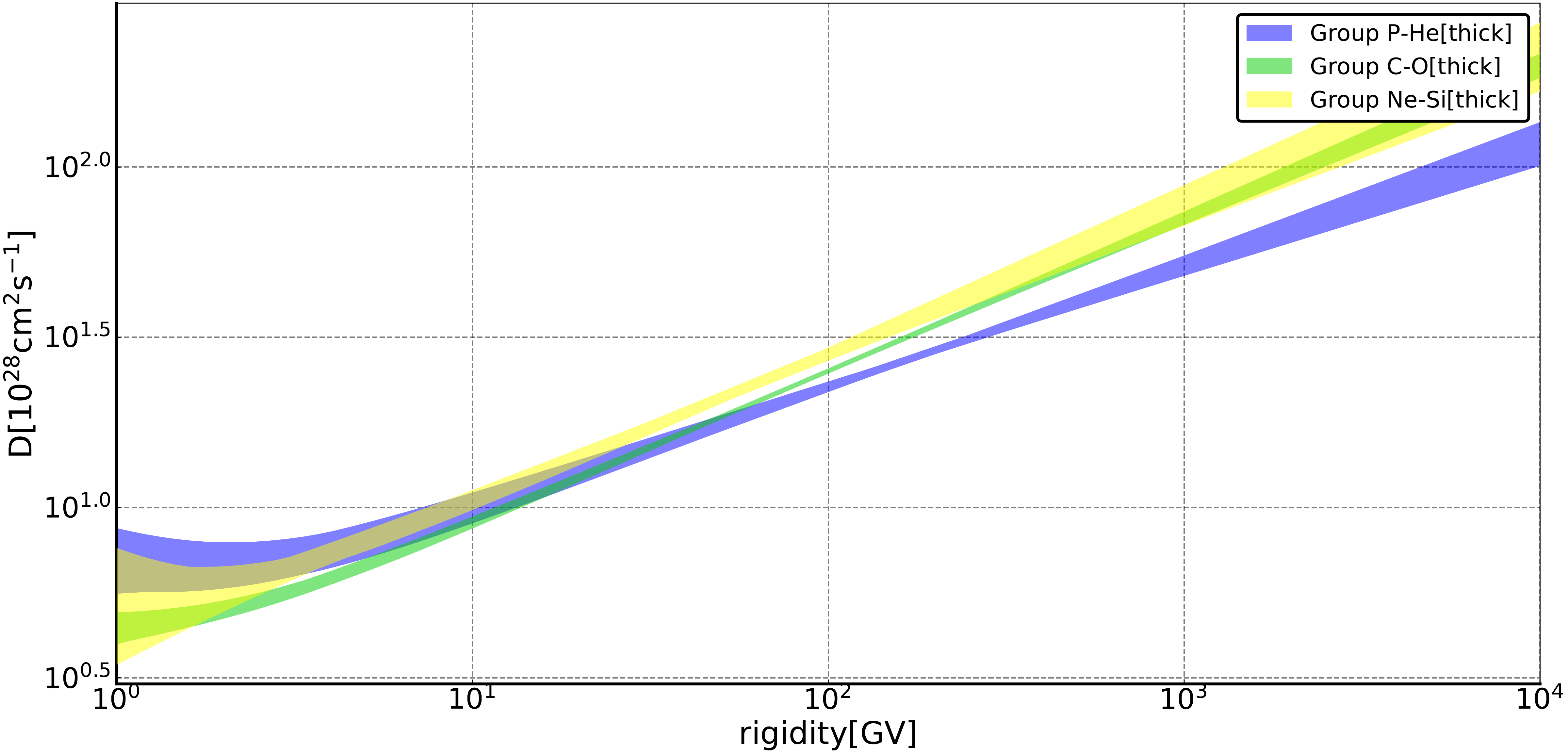}\\
\caption{\label{fig:diffusion} The diffuse coefficient D changes with rigidity R, compared between three nuclei groups: P-He (blue), C-O (green) and Ne-Si (yellow).
Top: Halo height h is fixed to 3.44 kpc inferred from combined B/C+$^{10}$Be/$^{9}$Be fit. Bottom: Halo height h is fixed to 7.17 kpc inferred from combined B/C+$^{10}$Be/$^{9}$Be+Be/B fit.}
\end{figure}
\begin{table*}
 \caption{ The best-fit values and posterior 95\% range of all parameters. Halo height are fixed to 3.44 kpc inferred from combined B/C+$^{10}$Be/$^{9}$Be fit.\label{tab:f}}
\begin{ruledtabular}
\begin{tabular}{ccccc}
 Parameter&P-He&C-O&Ne-Si\\ \hline
 $D_0(10^{28}~\rm{cm^2s^{-1}})$&3.586 [3.456,4.052]&3.190 [3.076,3.327]&3.805 [3.438,3.975]\\
 $\delta$&0.377 [0.350,0.386]&0.452 [0.439,0.463]&0.436 [0.417,0.469]\\
 $V_a$ (km/s)&21.847 [20.628,25.636]&21.154 [19.443,22.700]&23.413 [18.594,24.745]\\
 $\eta$&-0.574 [-0.764,-0.372]&-0.482 [-0.562,-0.305]&-0.406 [-0.911,-0.036]\\ 
 $\phi$ (GV)&0.554 [0.546,0.562]\footnote{the best-fit modulation potential for antiproton is 0.484~GV}&0.788 [0.775,0.805]&0.783 [0.715,0.823]\\ 
 \hline
 $\chi^2_{\rm{min}}/n_{\rm{d.o.f.}}$&167.41/212&142.37/228&249.25/281\\
\end{tabular}
\end{ruledtabular}
\end{table*}
\begin{table*}
 \caption{ The best-fit values and posterior 95\% range of all parameters. Halo height are fixed to 7.17 kpc inferred from combined B/C+$^{10}$Be/$^{9}$Be+Be/B fit.\label{tab:lf}}
\begin{ruledtabular}
\begin{tabular}{ccccc}
 Parameter&P-He&C-O&Ne-Si\\ \hline
 $D_0(10^{28}~\rm{cm^2s^{-1}})$&7.318 [6.633,8.085]&5.904 [5.746,6.198]&6.897 [6.554,7.557]\\
 $\delta$&0.350 [0.324,0.376]&0.449 [0.436,0.459]&0.439 [0.408,0.458]\\
 $V_a$ (km/s)&24.053 [21.268,27.477]&20.110 [19.146,22.515]&21.353 [19.455,25.425]\\
 $\eta$&-0.605 [-0.776,-0.390]&-0.442 [-0.561,-0.302]&-0.506 [-0.805,-0.00114]\\ 
 $\phi$ (GV)&0.545 [0.538,0.558]\footnote{the best-fit modulation potential for antiproton is 0.340~GV}&0.789 [0.774,0.803]&0.760 [0.724,0.822]\\ 
 \hline
 $\chi^2_{\rm{min}}/n_{\rm{d.o.f.}}$&175.51/212&143.96/228&246.03/281\\
\end{tabular}
\end{ruledtabular}
\end{table*}

In this section, we propose another possible solution for the F anomaly. By analyzing the relationship between the total reaction cross section and the mass number,
\cite{Johannesson:2016rlh} provided a rough estimation of the effective propagation distance of CR nuclei:
\begin{equation}
\label{eq:effective}
  	\left \langle x \right \rangle_A \sim 2.7~{\rm kpc}~ (\frac{A}{12})^{-1/3} (\frac{R}{R_0})^{\delta/2},
\end{equation}
where $R$ is the rigidity of CRs, $R_0=4~$GV is the reference rigidity, and $\delta$ is the slope index of diffusion coefficient. Since the propagation distance decreases with the increase of $A$, the average diffusion coefficient of CRs reaching Earth should vary with $A$ if the diffusion coefficient is spatially dependent.

We have introduced the spatially dependent diffusion model in a previous paper \cite{PhysRevD.104.123001}, where the spatial dependency of the diffusion coefficient traces the CR source distribution.
In the inner Galaxy, the distribution of the supernova remnants and the OB stars reach their maxima at a distance of $\sim4$ kpc from the Galactic center (see Fig.~1 in \cite{Ackermann:2012pya}). 
Considering that turbulence in the Galactic disk is mainly
generated by stellar feedback (such as the supernova explosions), the diffusion coefficient could be smaller at the position where the source distribution is more concentrated.
We analyze three nuclei groups with different $A$, namely the P-He group, C-O group, and Ne-Si group. According to Eq.~(\ref{eq:effective}), their effective propagation distance at 40 GV are: 8.9 kpc (P-He), 4.1 kpc (C-O), and 3.4 kpc (Ne-Si). The effective propagation area of the P-He group covers the most concentrated area of CR sources in the Galaxy, which means that their average diffusion coefficient is expected to be smaller than the other groups, while the average diffusion coefficient of the Ne-Si group is expected to be the largest. 

We perform a series of Bayesian fits for $\bar p/p$, B/C, and F/Si, respectively, to get the diffusion coefficients for the corresponding groups. The original GAL12 parametrization is adopted for the production cross section. The constrained diffusion coefficients are shown in Fig.~\ref{fig:diffusion}. As expected, the diffusion coefficient increases from the lighter nuclei to heavier nuclei for $R\gtrsim50$~GV. We check two different halo sizes (3.44 kpc and 7.17 kpc), since the AMS-02 Be/B data supports a larger halo height \cite{PhysRevD.104.123001,Evoli:2019iih,DeLaTorreLuque:2021yfq,Weinrich:2020ftb} than that inferred from $^{10}$Be/$^{9}$Be.
In Table~\ref{tab:f} and Table~\ref{tab:lf}, we list the best-fit values of transport parameters with posterior 95\% confidence range.
The posterior confidence intervals are slightly larger for the P-He and Ne-Si group than for the C-O group, which is mainly attributed to the larger experimental uncertainties of the $\bar p$ and F fluxes.

With the help of the Voyager measurements of un-modulated fluxes \cite{Cummings:2016pdr}, we obtain the modulation potentials for different groups. For P-He group we use $\bar p$, proton, and he fluxes \cite{Aguilar:2021tos} taken during May 2011-May 2018, which is the same period for C-O group where we used B, C, N, and O fluxes \cite{Aguilar:2021tos}. While for Ne-Si group we use Ne, Mg, and Si fluxes \cite{AMS:2020cai} taken during May 2011-May 2018, together with F flux \cite{AMS:2021tnd} taken during May 2011-Oct 2019.
The obtained $\phi$ of the three groups should be similar since these measurements were taken during a similar period.
As shown in Table~\ref{tab:f} and Table~\ref{tab:lf}, the C-O and Ne-Si groups have similar $\phi$ of $\sim0.78$ GV. However, the P-He group predict a smaller potential $\phi\sim0.55$ GV, which is also found in \cite{Yuan:2018vgk}. The difference may be attributed to the oversimplification of the force field approximation \cite{Wang:2019xtu}.

We notice that the C-O and Ne-Si groups predict a Iroshnikov-Kraichnan slope ($\delta\sim1/2$), while the P-He group predict a Kolmogorov-like slope ($\delta\sim1/3$). 
The production mechanism of secondary nuclei like B is significantly different from that of antiproton. The former keeps the energy per nucleon of secondary particles the same as that of the primary particles, while the latter is the convolution of the primary
spectra and the differential cross section \cite{diMauro:2014zea}. It may cause the difference in slope. An alternative explanation of the difference is that $\delta$ is also spatially dependent \cite{Fermi-LAT:2016zaq,Yang:2016jda}.

The posterior propagation parameters of C-O group and Ne-Si group are quite similar, except for the reference diffusion coefficient $D_0$ (defined at 4 GV). The $D_0$ of C-O group is smaller than that of Ne-Si group. This feature implies that F is mostly (or purely) of secondary origin produced in a faster diffusion area. The fitting result prefers the prediction from the effective propagation distance and the spatially dependent diffusion model.

\section{SUMMARY\label{sec:conclusion}}
Considering the production cross-section uncertainties of the secondary CRs, we test whether the B/C and F/Si measured by AMS-02 can be consistently interpreted by the same framework of CR propagation. We adopt four different parametrizations of cross section to discuss the systematic error. For each parametrization, the normalizations of the relevant cross sections are constrained by the latest measurements, which is a major difference from the previous works. 

We first perform a preliminary fit to the cross-section data to obtain the mean and variance for the cross-section normalization of each channel. To illustrate how the cross-section uncertainties could influence the F/Si ratio, we assume the propagation parameters inferred from the B/C data and draw the predicted F/Si with the 68\% and 95\% bands according to the data restriction. The result shows a significant excess to the measured F/Si even considering the error bands. Then based on the Bayesian inference, we perform combined fits to the B/C, F/Si, and cross-section normalizations obtained in the preliminary step. Among the four cross-section models, the $\chi^2_{\rm{min}}/n_{\rm{d.o.f.}}$ obtained for the GAL12 and WE93 parametrizations are close to 1, implying a globally good fit. In contrast, the goodness of fits for the GAL22 and TS00 cases are very poor and are excluded with confidence levels of more than $99.9\%$. The main reason for the difference is that the latter models predict significantly higher secondary contributions from nuclei with $A>36$, which aggravates the overestimation of F. We should emphasize that although the overall goodness of fits for the GAL12 and WE93 models are acceptable, the exclusive goodness of fits of the cross-section data are very poor with $\chi^2_{\rm{cs}}/n_{\rm{cs}}\gtrsim2$. The total $\chi^2$ statistic is dominated by the CR part, so systematic deviations between the best-fit cross-section normalizations and the measurements are required to ensure a better overall $\chi^2$. We find that all the best-fit normalizations for the important channels for  F production are significantly smaller than the experimental values, while the opposite is true for the case of B. This indicates that the F anomaly can hardly be interpreted by neither the random errors of the cross-section measurements nor the systematic errors induced by the differences between the existing cross-section models.

The available cross-section data are mainly below 1 GeV$/n$, and the secondary production for higher energies is usually given by the extrapolation based on different formulae. If the cross sections for channels producing F decrease rapidly in 1-10 GeV$/n$, the F flux could be suppressed compared with the flux calculated by the current cross-section models. Future cross-section measurements at higher energies will test this possibility. Alternatively, smaller cross sections for the channels producing F from heavy nuclei ($A>28$) may help to suppress the F flux. However, although most of these channels have not been studied, the measured cross section of $^{56}\text{Fe} \longrightarrow ^{19}\text{F}$ indicates that the current WE93 model is already smaller than the data, which may not support further smaller cross sections.

Meanwhile, we introduce the spatially dependent diffusion model as an alternative interpretation of the F anomaly. The effective propagation distance of CR nuclei decreases with the mass number, which makes the average diffusion coefficient of the Ne-Si group larger than that of the C-O group under the spatially dependent diffusion. Thus, the overestimation of F/Si could be solved owing to the larger diffusion coefficient, and the B/C and F/Si data could be explained consistently. We perform a series of fits to the AMS-02 $\bar p/p$, B/C, and F/Si data to get the average diffusion coefficients of these three different nuclei groups. The results indicate that the needed average diffusion coefficient increases with the nuclei mass number. This tendency is consistent with the prediction of the spatially dependent diffusion.

 
\acknowledgments
This work is supported by the National Natural Science Foundation of China under Grants No. 12175248, No. 12105292, and No. U1738209.

\bibliography{apssamp}

\appendix

\section{Plots of Cross Section Data\label{app:xsdata}}
Here we show the plots of the most important channels needed for analyzing B and F production in the paper.
Only collisions with the hydrogen target are presented for simplicity. As implemented in {\footnotesize GALPROP}'s fragmentation routine, the collisions with the helium target are calculated using a parametrization by Ferrando \cite{Ferrando:1988tw} where the interstellar gas ratio of helium to hydrogen is set to be 0.11.

The available measurements are mostly based on the {\footnotesize GALPROP} cross-section data (labeled as [GAL]) assembled in the file \texttt{isotope\_cs.dat}, others from Zeitlin \cite{Zeitlin:2001ye,Zeitlin:2007sm,Zeitlin:2011qg} (labeled as [Ze01]/[Ze06/[Ze11]]), Villagrasa-Canton \cite{Villagrasa-Canton:2006exk} (labeled as [Vi07]), Flesch \cite{FLESCH2001237} (labeled as [Fl01]), Napolitani \cite{Napolitani:2004fw} (labeled as [Na04]) and EXFOR database (labeled as [EXFOR]).

Together with the data we also draw the GAL12, GAL22, WE93, and TS00 parametrizations, which were taken from the {\footnotesize GALPROP} code that haven't been re-normalized in Sec.~\ref{sec:result}.
As designed in the evaluation routines of GAL12 and GAL22,
the contributions of ghost nuclei $^{10}\text{C}$ and $^{11}\text{C}$ are directly counted as the cumulative $^{10}\text{B}$ and $^{11}\text{B}$ if the projectile is $\rm{^{16}O}$ or $\rm{^{14}N}$. To compare GAL12 and GAL22 with other parametrizations and also the data, we added the cross section of ghost nuclei together with the corresponding stable nuclei.
This is significantly different from how Genolini \cite{Genolini:2018ekk} treated ghost nuclei, as they subtract ghost nuclei based on Webber's prediction \cite{Webber:1998ex,Webber_2003} of the proportion to the cumulative cross section, rather than adding them together. While the final result still takes the contributions from ghost nuclei into consideration.

The formula of isotopic cross sections for projectile nuclei ($Z_i,A_i$) into fragment nuclei ($Z_f,A_f$) has been constructed by Webber \cite{Webber_2003}:
\begin{equation}
\label{eq:webber-for}
 f_1(Z_f,A_f,Z_i,A_i)=\frac{1}{\delta_{Z_f}\sqrt{2\pi}}{\rm exp}\frac{-(N-N_{Z_f})^2}{2\delta_{Z_f}^2},
\end{equation}
where $\delta_{Z_f}$ is the characteristic width of the individual mass distributions of charge $Z_f$, and $N_{Z_f}$ is the neutron excess of the mass centroid.
Although the formula of isotopic cross sections is independent of energy, the values of $\delta_{Z_f}$ and $N_{Z_f}$ were evaluated according to the data available at 500-600 MeV$/n$, which may be biased when estimating data at energies far beyond the region.
It is also worthy of notice that, data points from Zeitlin \cite{Zeitlin:2001ye,Zeitlin:2007sm,Zeitlin:2011qg} and Flesch \cite{FLESCH2001237} have been rescaled based on Webber's prediction \cite{Webber:1998ex,Webber_2003}.
The main reason is that these data are elemental charge changing cross sections describing the \textit{total} fragment of F or Ne, not specified isotopes.
To predict how the cross section is divided among the isotopes, we used Webber's parametrizations as a reference since the mass distributions of the fragments for individual
charges have been accurately predicted by their formula.

\begin{figure*}[htbp]
\label{channel1}
\caption{Channels: $X + p \longrightarrow ^{19}\text{F}$ and $X + p \longrightarrow ^{19}\text{Ne}$. The parametrizations shown in the figures were taken from the {\footnotesize GALPROP} code. In this work, we renormalized the parametrizations to fit the new data shown in the figures.}
    \centering
      \subfigure{
        \includegraphics[width=0.45\textwidth]{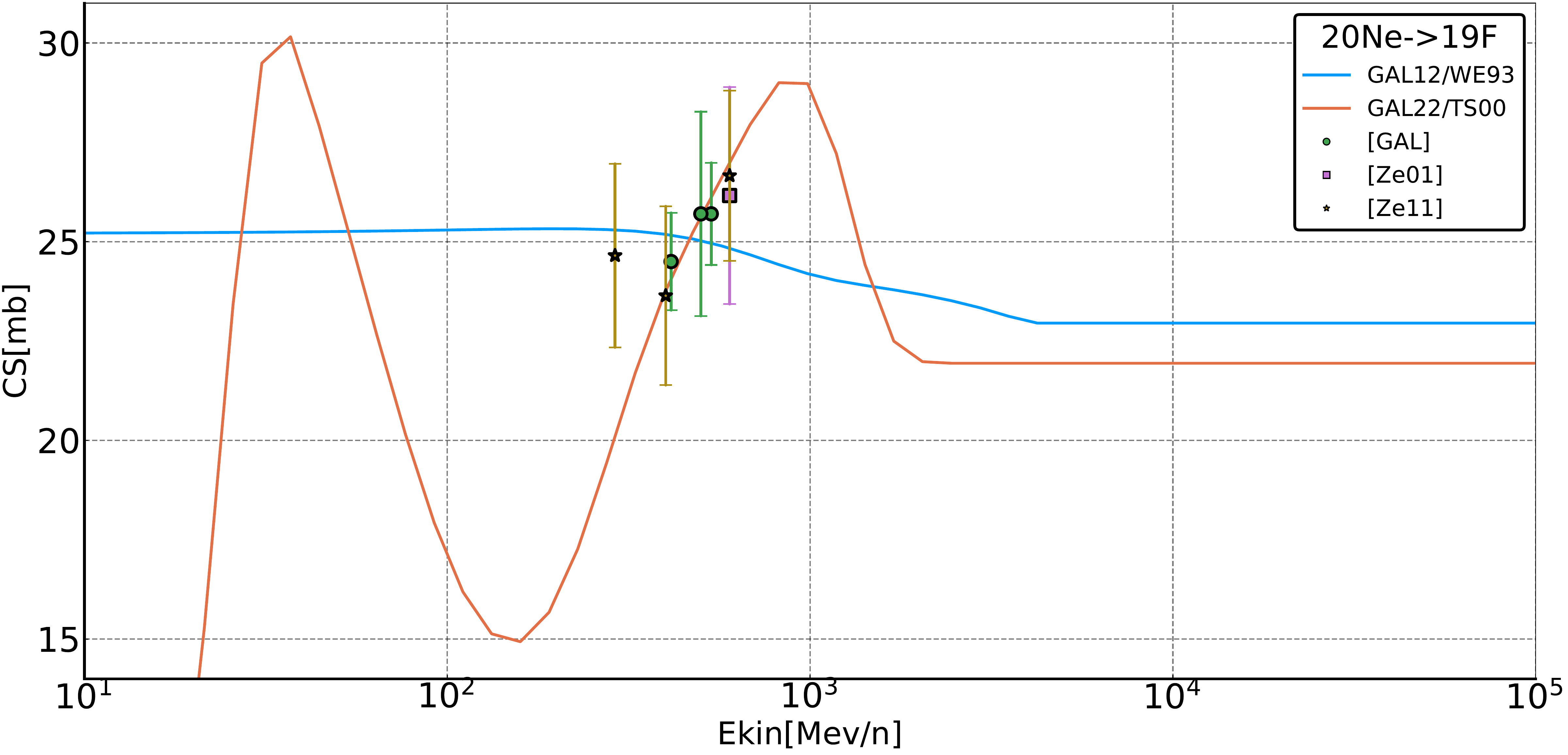}
      }
      \subfigure{
        \includegraphics[width=0.45\textwidth]{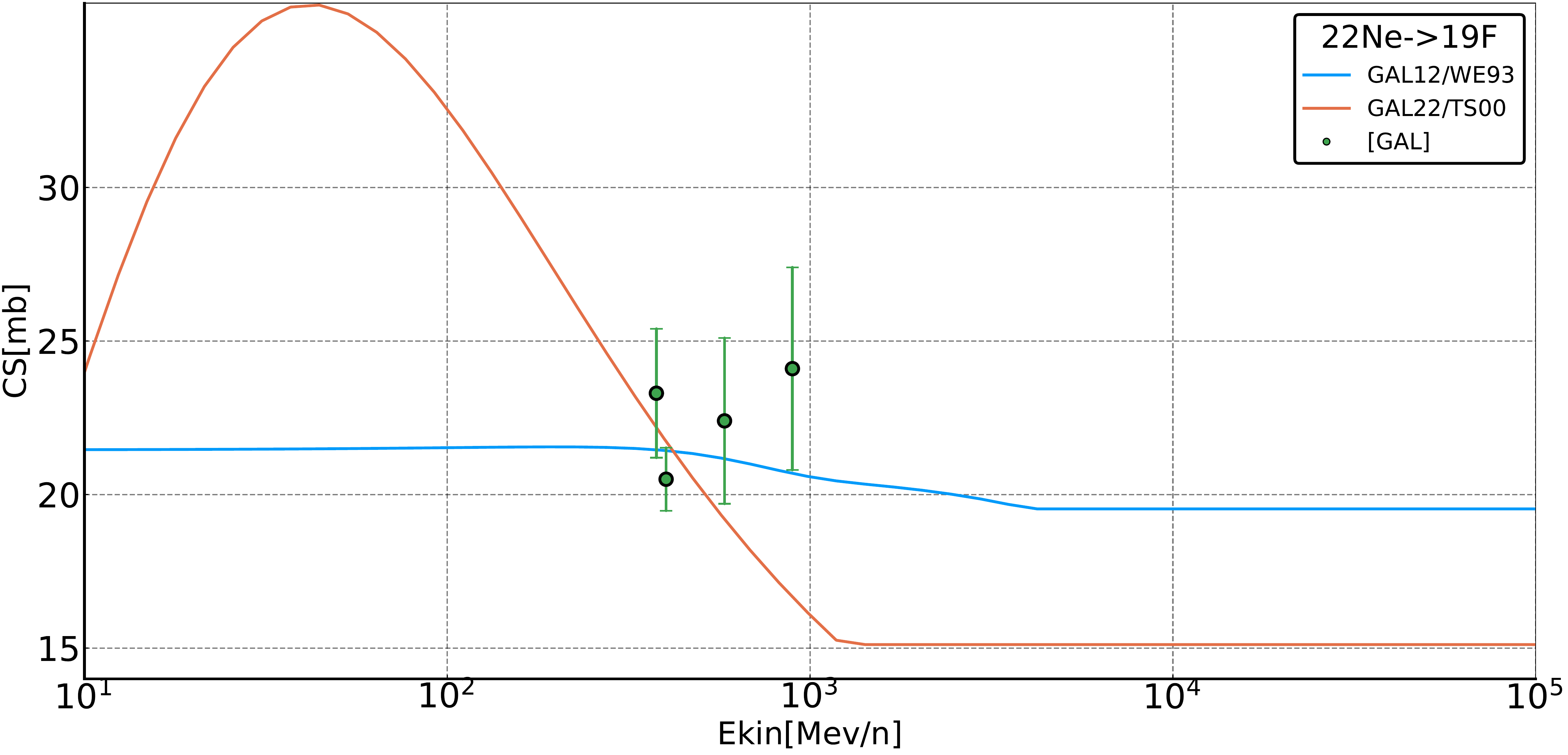}
      } \\
      \subfigure{
           \includegraphics[width=0.45\textwidth]{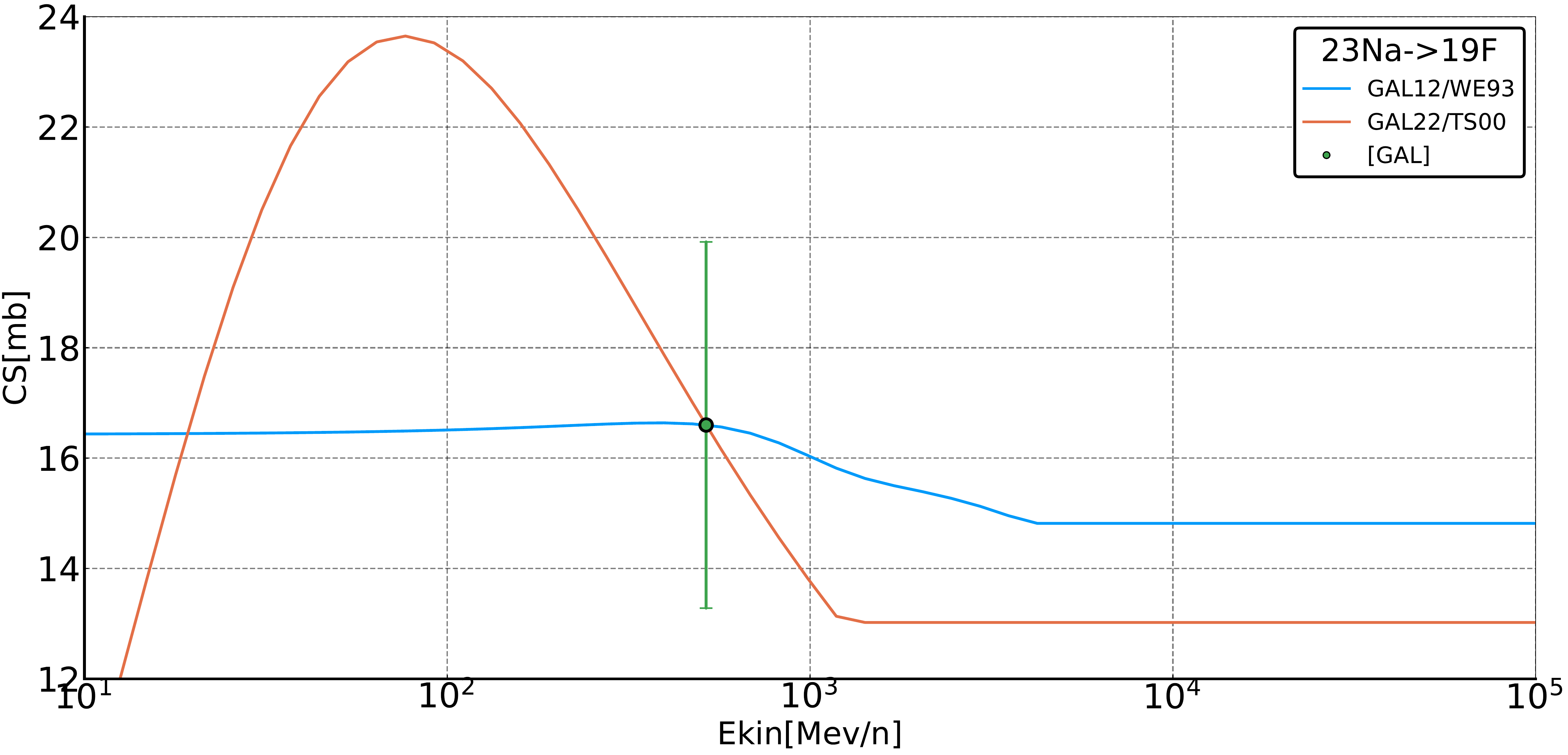}}
       \subfigure{    
           \includegraphics[width=0.45\textwidth]{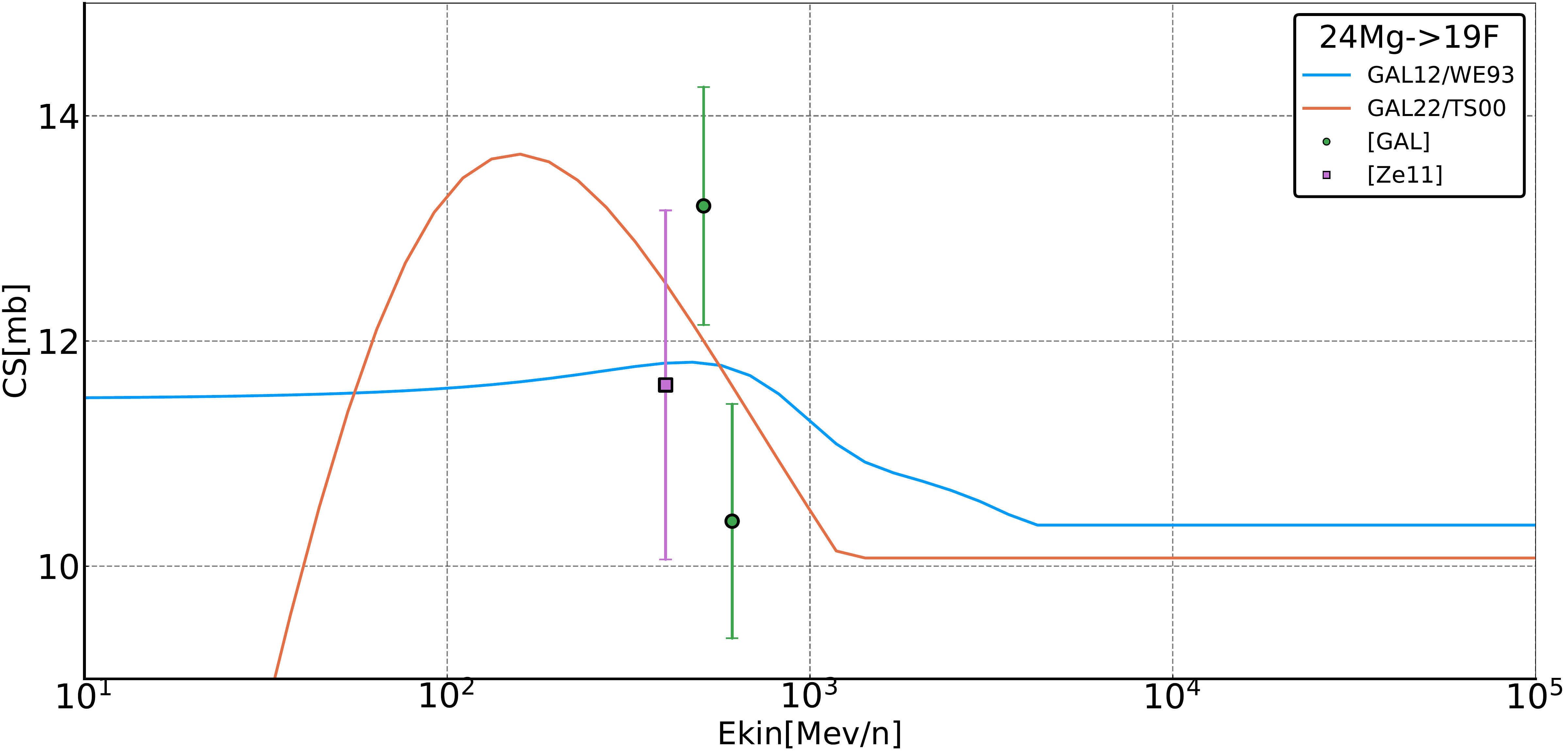}
      }\\
 
      \subfigure{
         \includegraphics[width=0.45\textwidth]{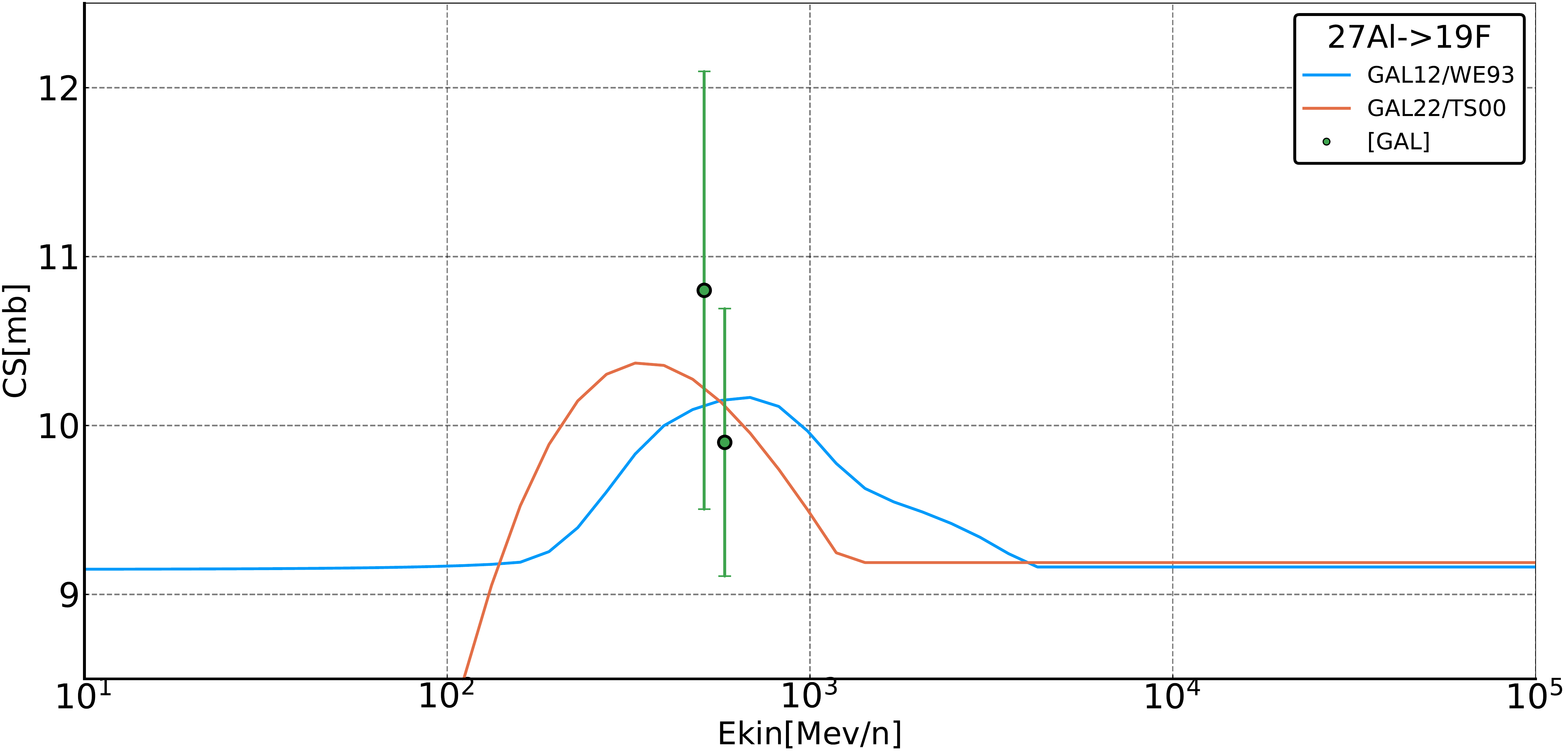}}
        \subfigure{  
           \includegraphics[width=0.45\textwidth]{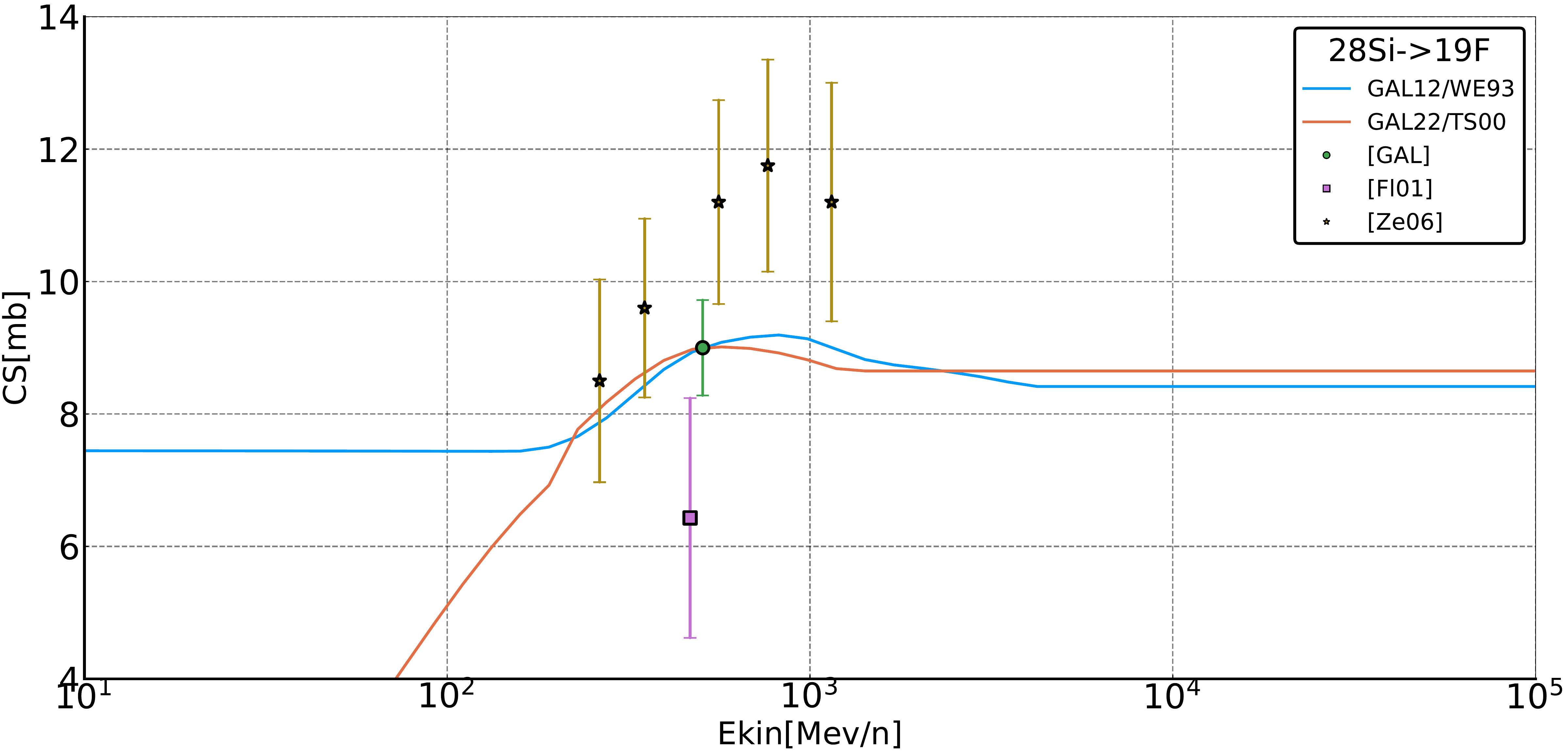}
      }\\
      \subfigure{
           \includegraphics[width=0.45\textwidth]{7-25-07.pdf}}
     \subfigure{
           \includegraphics[width=0.45\textwidth]{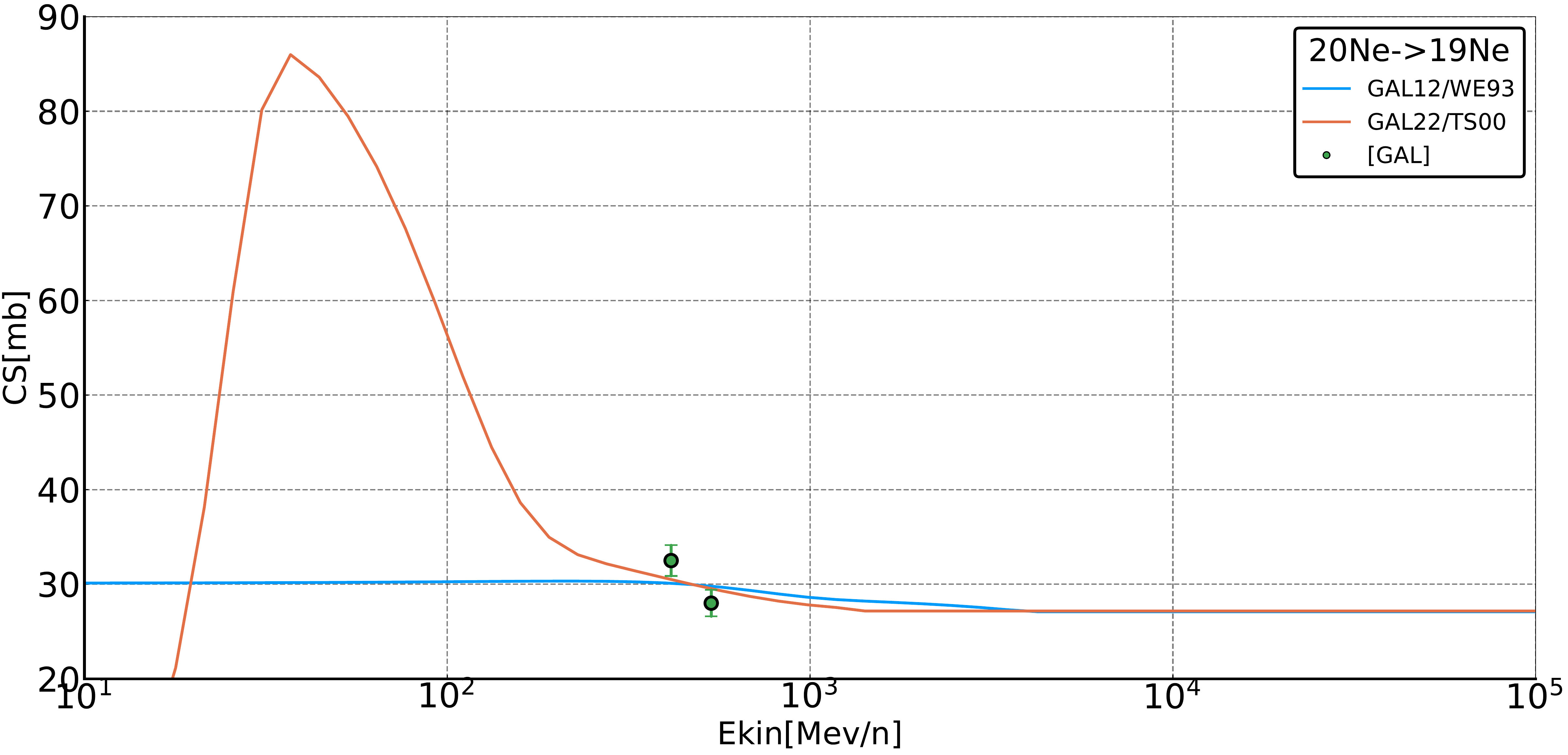}
      }\\
        
      \subfigure{
           \includegraphics[width=0.45\textwidth]{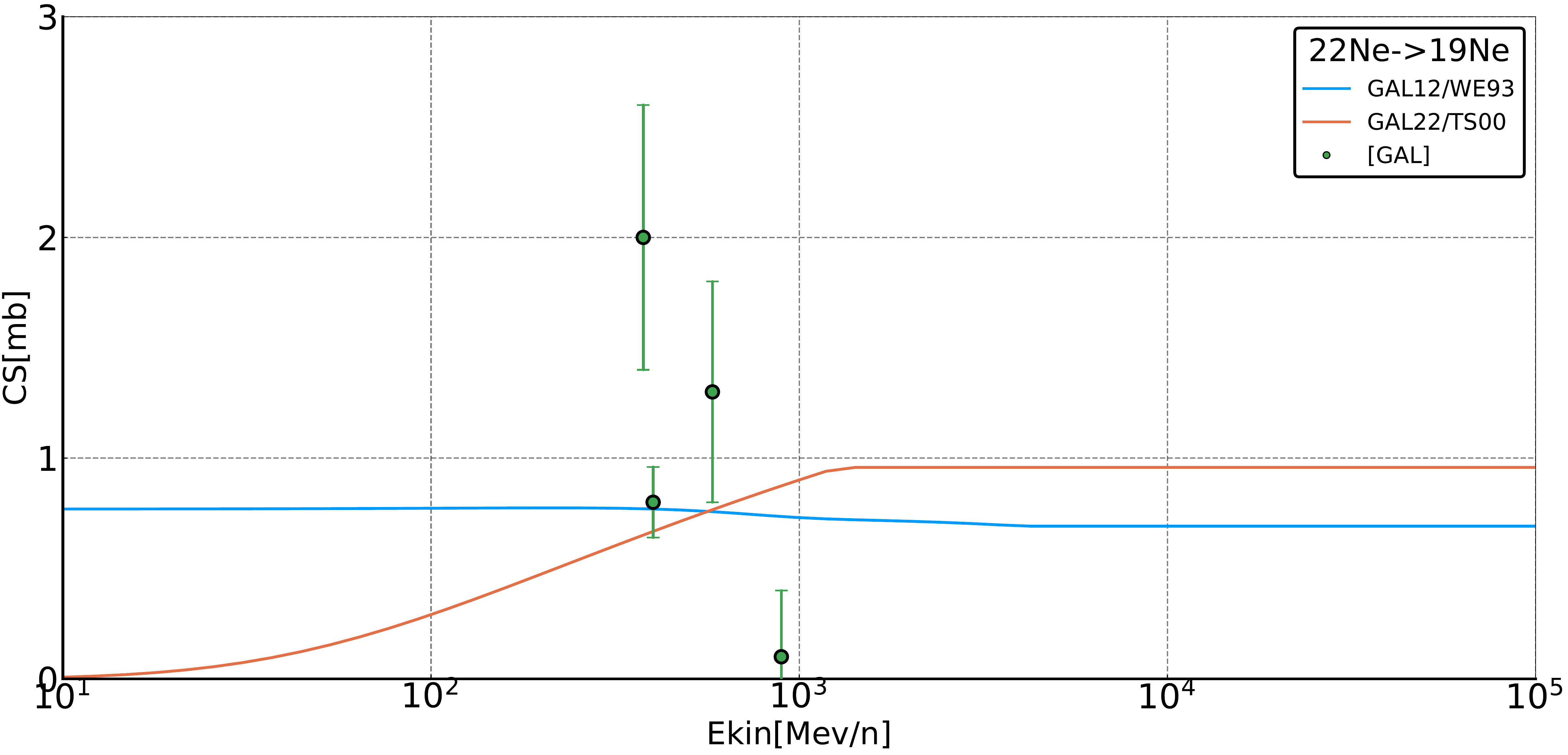}}
     \subfigure{
           \includegraphics[width=0.45\textwidth]{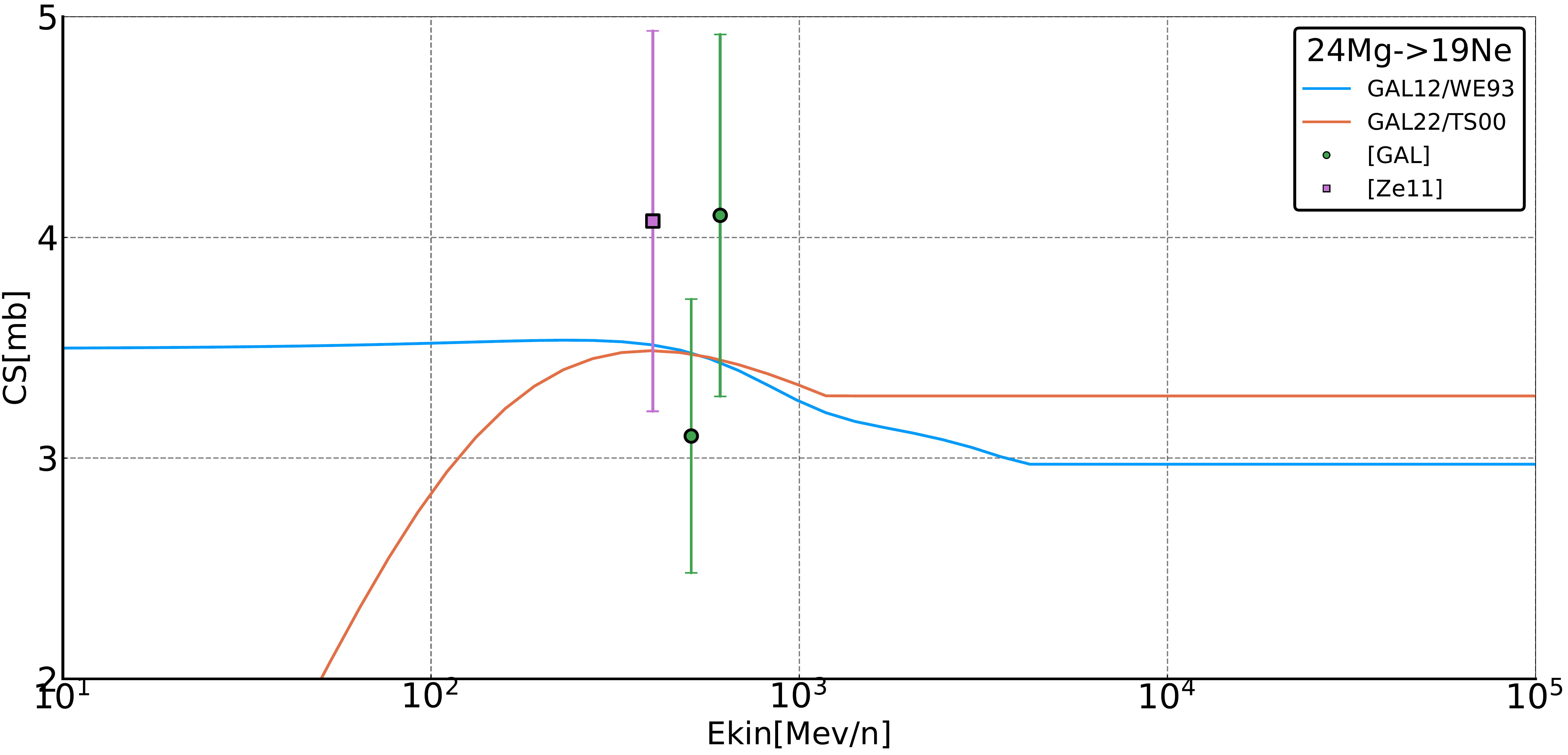}
      }
\end{figure*}

\begin{figure*}[htbp]
     \includegraphics[width=0.45\textwidth]{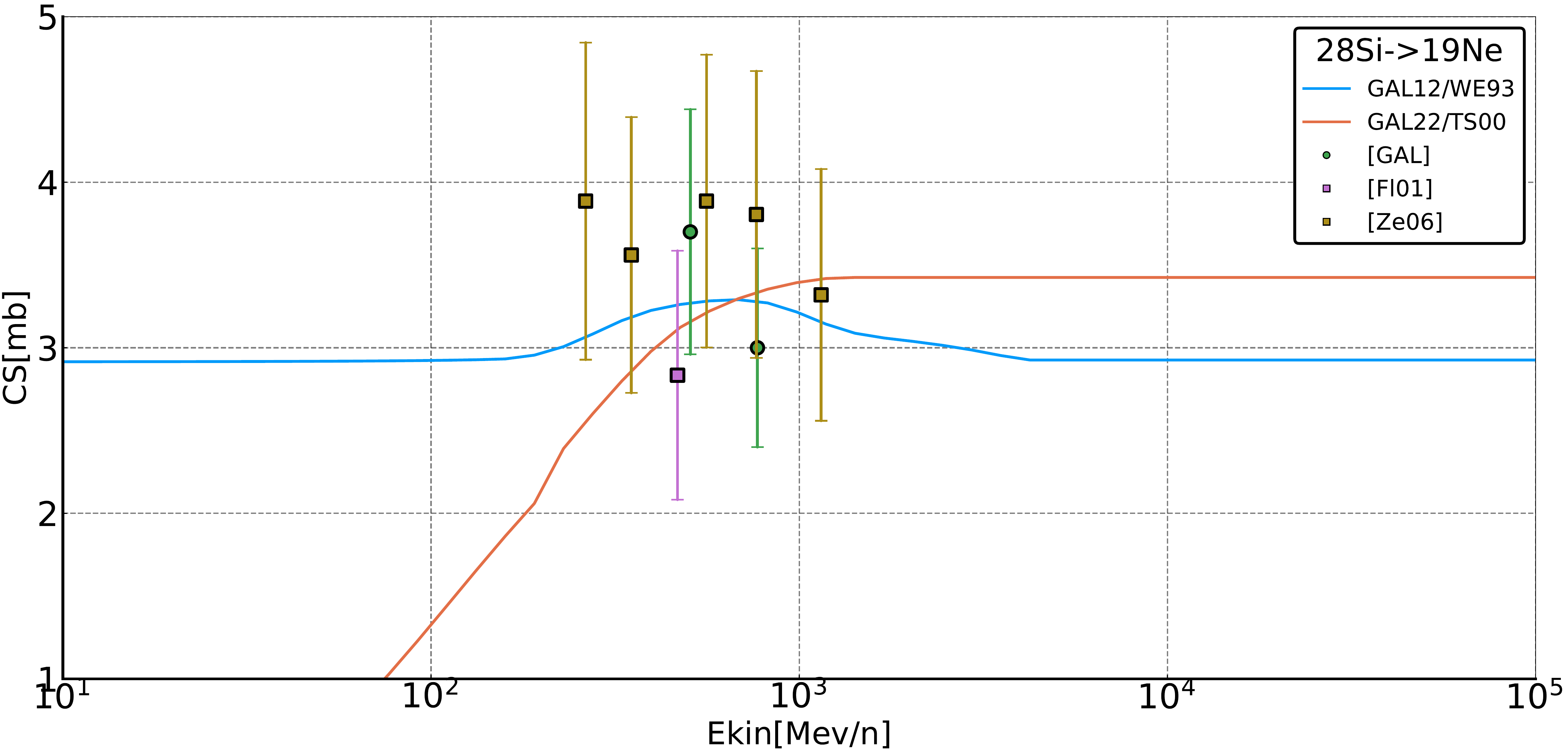}\\
\caption{Channels: $X + p \longrightarrow ^{10}\text{B}~(^{10}\text{C})$ and $X + p \longrightarrow ^{11}\text{B}~(^{11}\text{C})$. The parametrizations shown in the figures were taken from the {\footnotesize GALPROP} code. In this work, we renormalized the parametrizations to fit the new data shown in the figures.}
 \centering
      \subfigure{
           \includegraphics[width=0.45\textwidth]{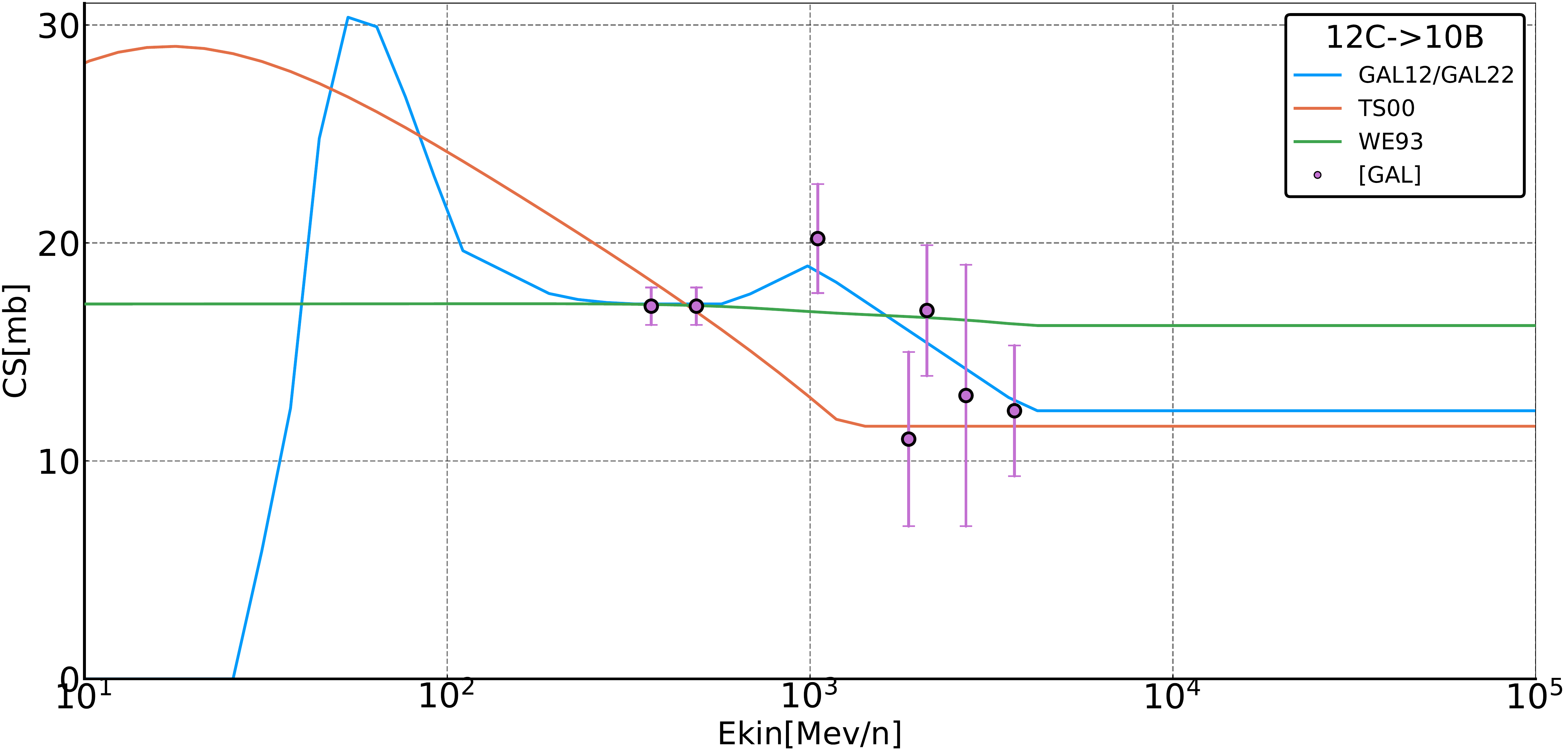}}
      \subfigure{
           \includegraphics[width=0.45\textwidth]{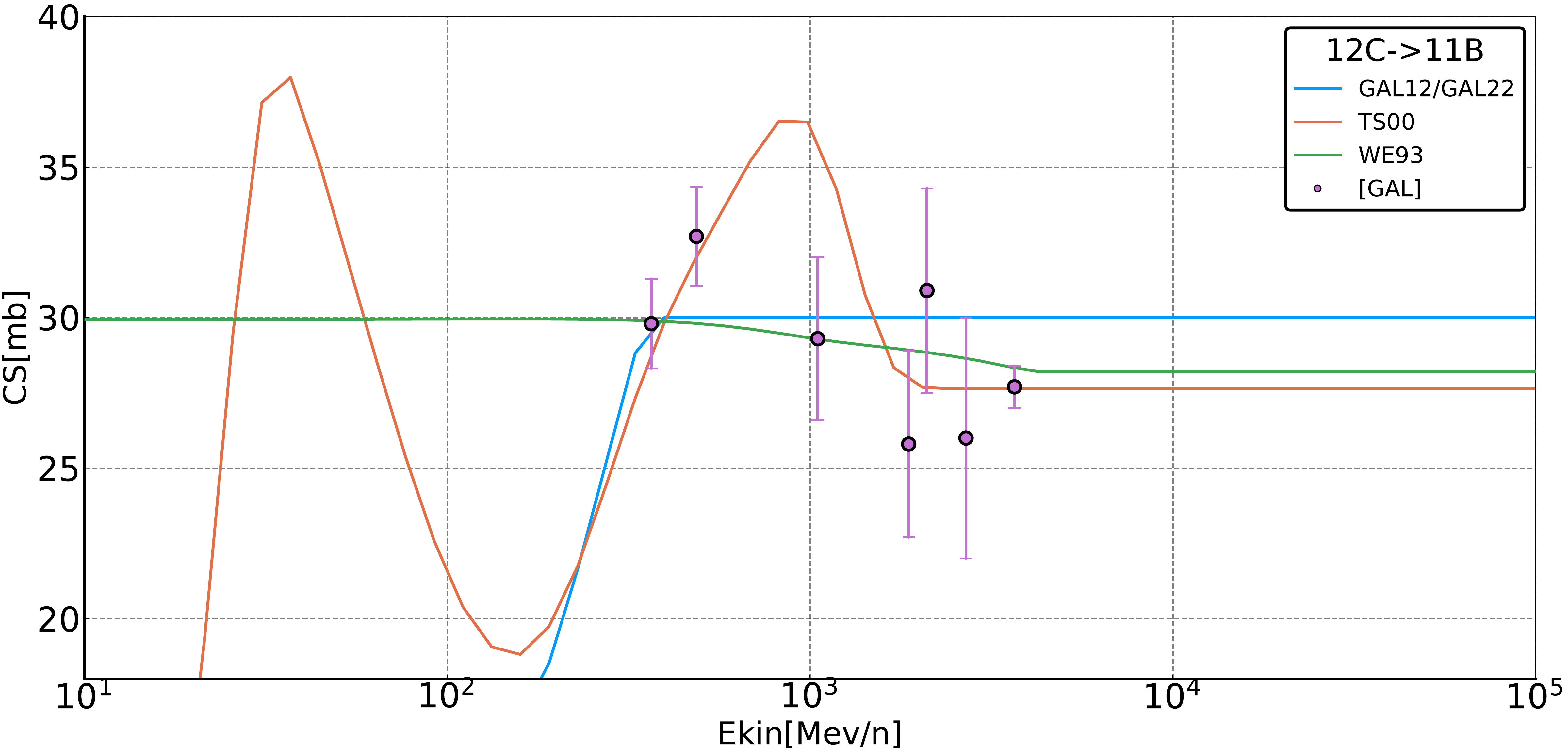}
      }\\
      \subfigure{
           \includegraphics[width=0.45\textwidth]{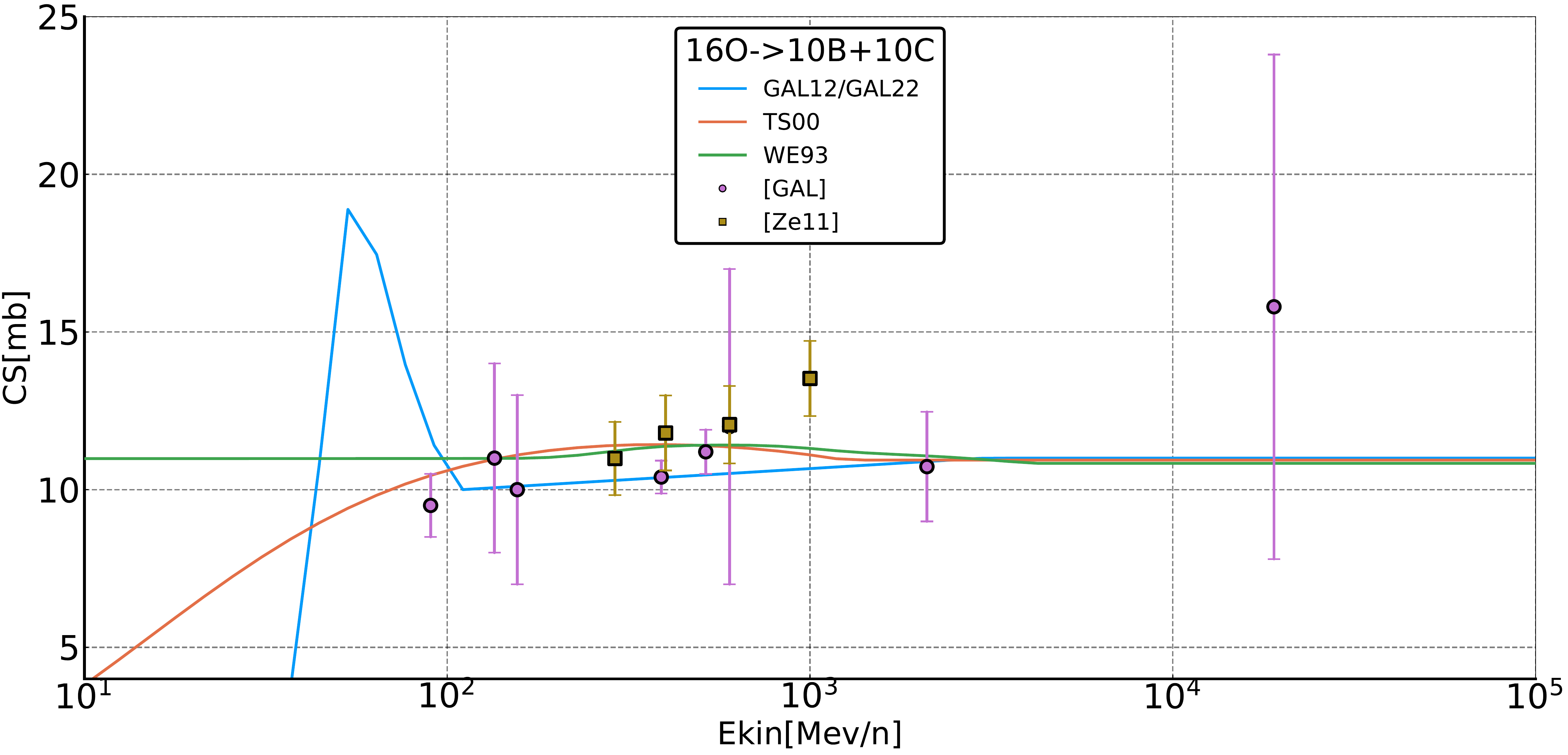}}
    \subfigure{
           \includegraphics[width=0.45\textwidth]{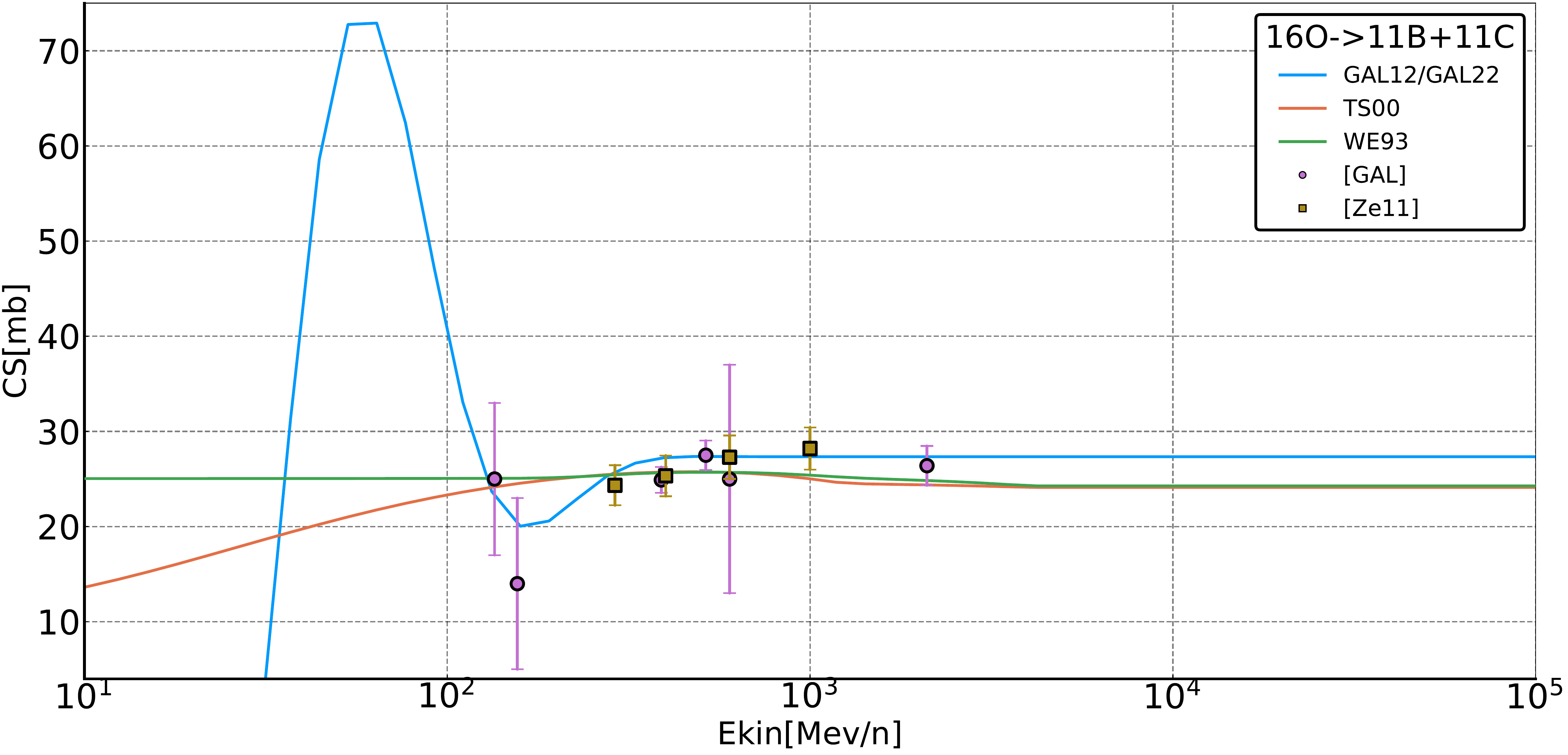}
      }\\
      \subfigure{
           \includegraphics[width=0.45\textwidth]{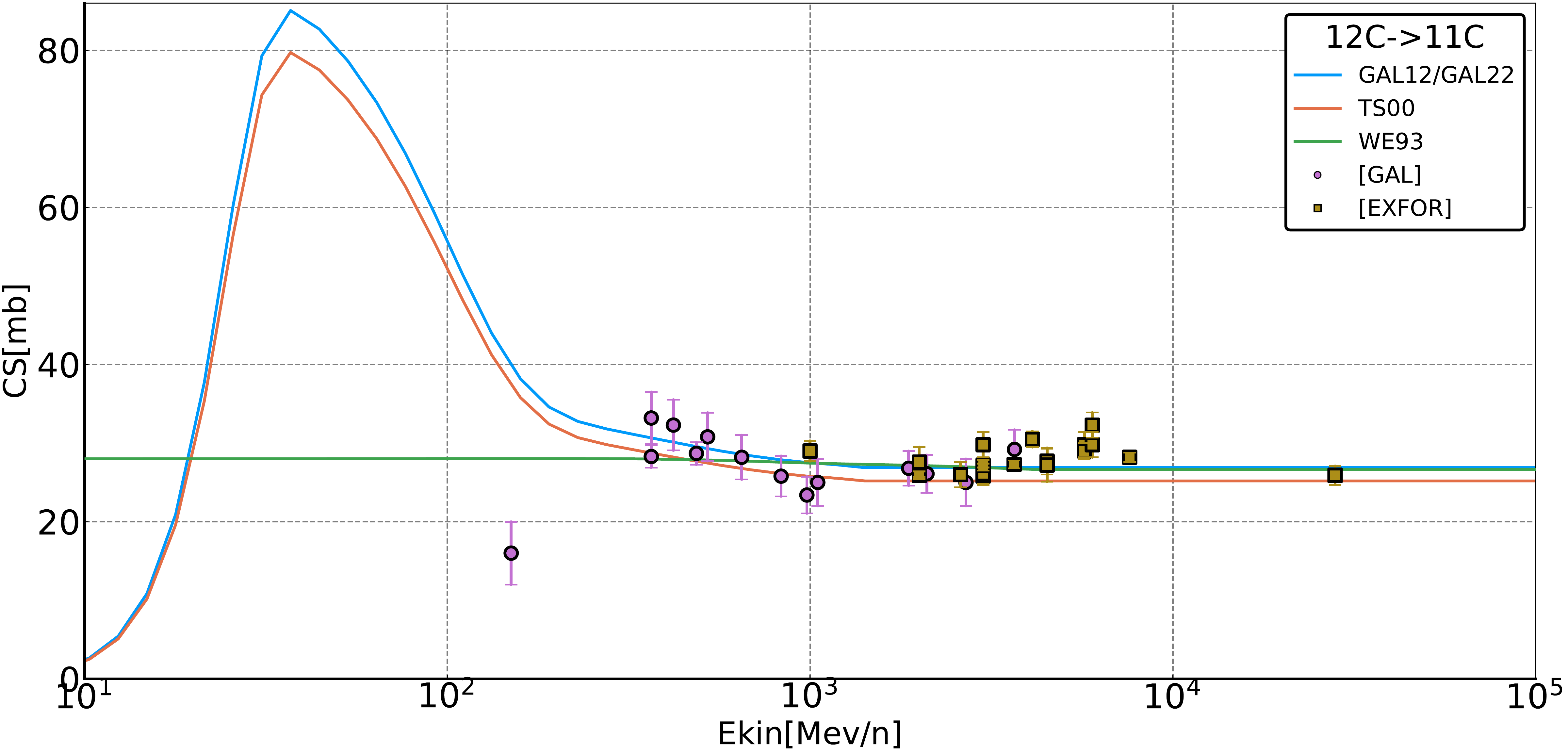}}
       \subfigure{
           \includegraphics[width=0.45\textwidth]{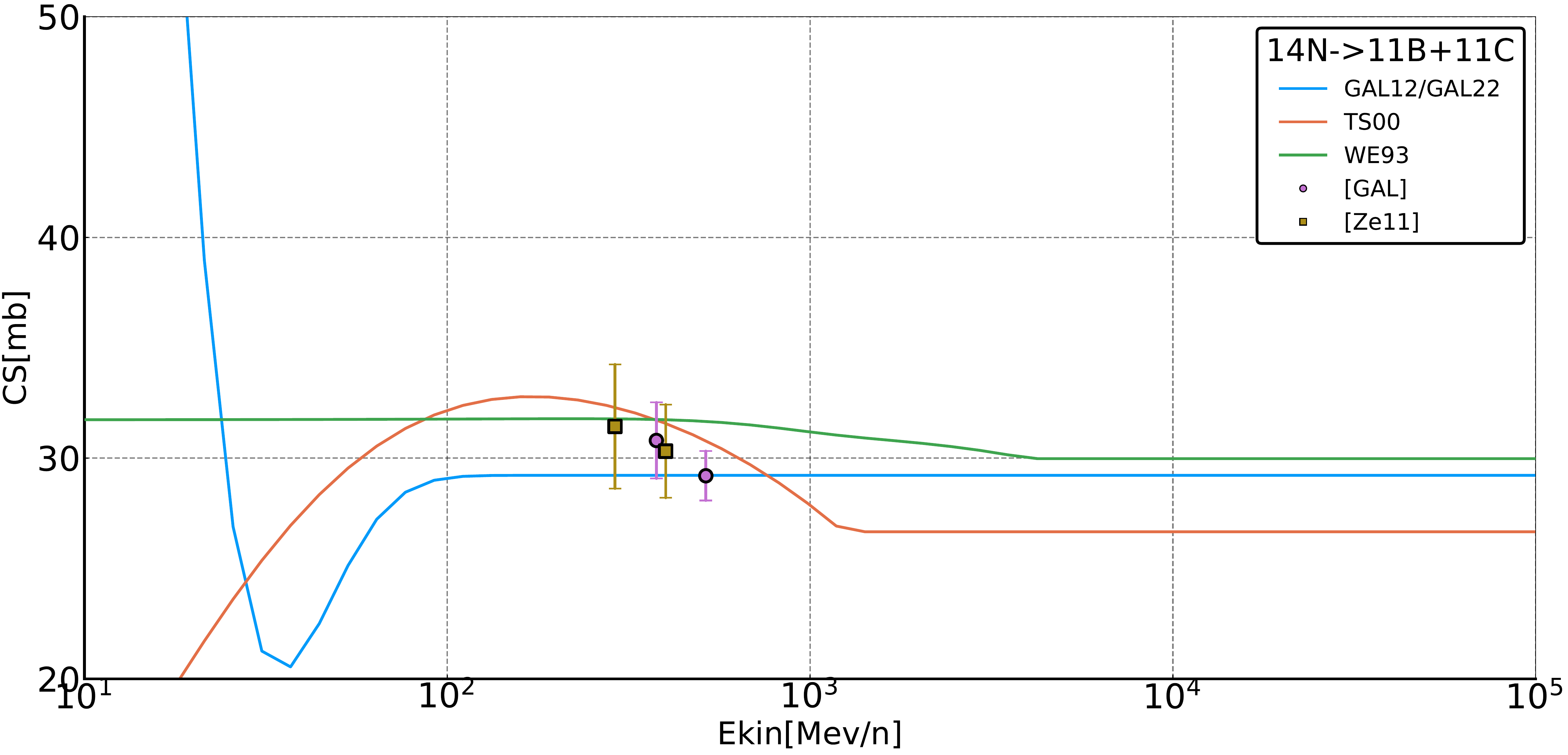}
      }\\
   \includegraphics[width=0.45\textwidth]{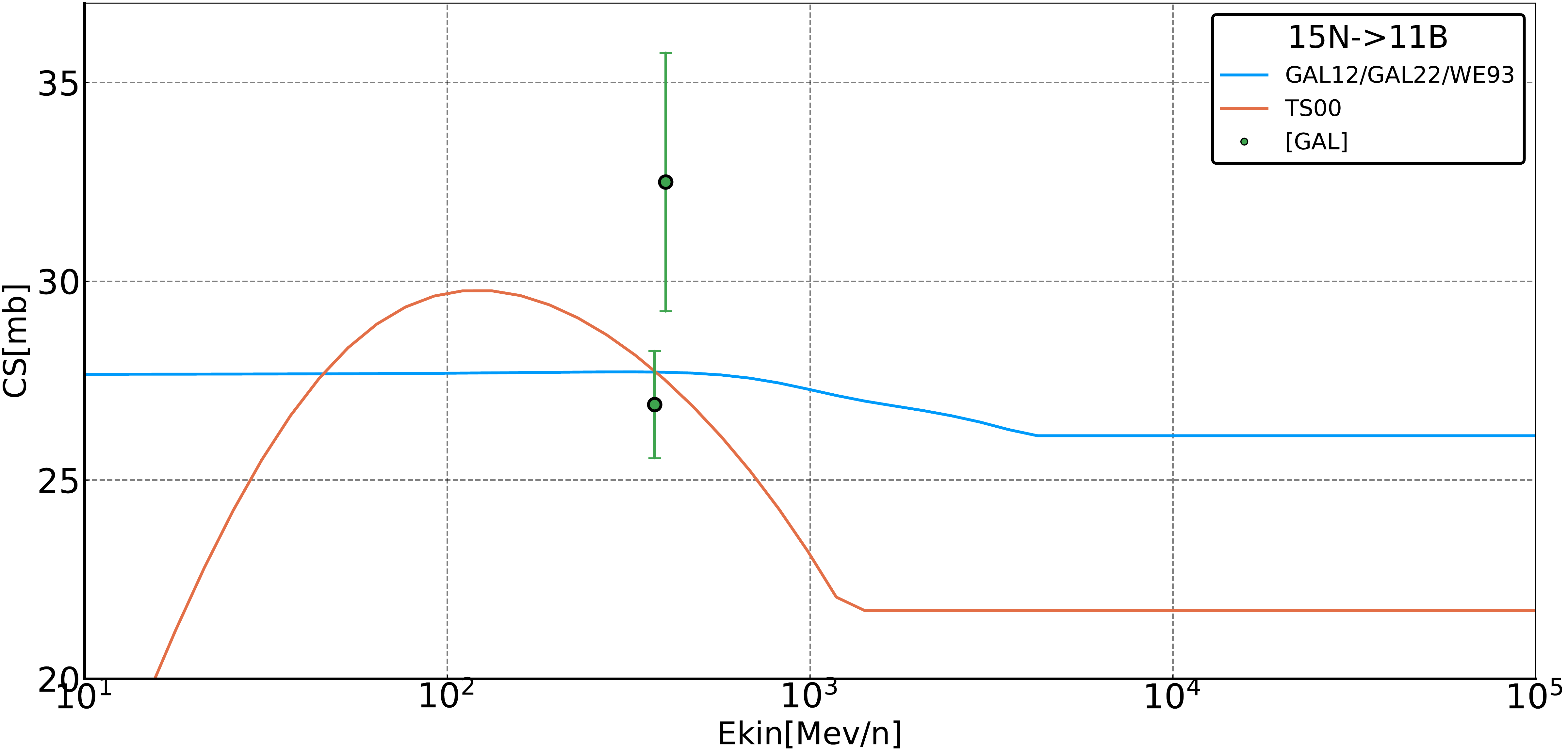}
\end{figure*}

\end{document}